\RequirePackage{ifpdf}
\documentclass{JHEP3}
\pdfoutput=1
\usepackage{graphics}
\usepackage{graphicx}
\usepackage{amsmath}
\usepackage[normalem]{ulem}
\usepackage{cite}
\usepackage{dsfont}
\usepackage{color}
\let\normalcolor\relax
\newcommand{\pFq}[5]{\, \! _{#1} F_{#2}\left( #3 \, ; \, #4 \, ; \, #5 \right)}

\newcommand{\epcolor}[1]{\textcolor{red}{#1}}

\def\be{\begin{equation}}
\def\ee{\end{equation}}
\def\bea{\begin{eqnarray}}
\def\eea{\end{eqnarray}}
\def\eps{\epsilon}
\def\nnb{\nonumber}
\def\ep{\epsilon}
\def\eps{\epsilon}
\def\dps{\displaystyle}

\preprint{
SI-HEP-2015-09\\
QFET-2015-10}
\title{Two-loop master integrals for non-leptonic heavy-to-heavy decays}

\author{Tobias Huber and Susanne Kr\"ankl\\
Theoretische Physik 1, Naturwissenschaftlich-Technische Fakult\"at,\\
Universit\"at Siegen, Walter-Flex-Stra{\ss}e 3, D-57068 Siegen, Germany\\

\email{huber@physik.uni-siegen.de,\\
       kraenkl@physik.uni-siegen.de}}

\abstract{We compute the two-loop master integrals for non-leptonic heavy-to-heavy decays analytically in a recently-proposed canonical basis. For this genuine two-loop, two-scale problem we first derive a basis for the master integrals that disentangles the kinematics from the space-time dimension in the differential equations, and subsequently solve the latter in terms of iterated integrals up to weight four.
The solution constitutes another valuable example of the finding of a canonical basis for two-loop master integrals that have two different internal masses,
and assumes a form that is ideally suited for a subsequent convolution with the light-cone distribution amplitude in the framework of QCD factorisation.}

\keywords{B-Physics, NLO computations, QCD, Heavy Quark Physics}

\begin{document}


\section{Introduction}
\label{sec:intro}

Non-leptonic $B$-decays are interesting for a number of phenomenological applications like the extraction of CKM elements and the study of CP asymmetries.
Their study has already entered the area of precision physics, both on the experimental~\cite{LHCbupgrate} and on the theoretical side. However, their theoretical description is complicated by the purely hadronic environment, entailing QCD effects from many widely separated scales. The two main approaches to non-leptonic $B$-decays are flavour symmetries of the light quarks~\cite{Zeppenfeld:1980ex} and factorisation frameworks such as pQCD~\cite{Keum:2000ph} and QCD factorisation (QCDF)~\cite{Beneke:1999br,Beneke:2000ry,Beneke:2001ev}. In the latter framework, next-to-leading order (NLO) corrections to both, heavy-to-heavy~\cite{Beneke:2000ry} and heavy-to-light~\cite{Beneke:1999br,Beneke:2003zv} transitions have been known since more than a decade. More recently, also next-to-next-to-leading order (NNLO) results for heavy-to-light decays have become available~\cite{Bell:2007tv,Bell:2009nk,Beneke:2009ek,Bell:2014zya,BBHL}.
In the present article, we consider NNLO corrections also to the heavy-to-heavy decays such as $B\rightarrow D\pi$ in the framework of QCDF~\cite{Huber:2014kaa}.
In the heavy-quark limit, the decay amplitude for $\bar B^0 \rightarrow D^{+}\pi^{-}$ is given by~\cite{Beneke:2000ry}
\begin{align}
 \langle D^{+}\pi^{-}|\mathcal{O}_i |\bar{B}^0 \rangle = \sum_j F_j^{B\rightarrow D} (m_\pi^2) \int_0^1 du \,T_{ij}(u) \Phi_{\pi}(u) \, ,  \label{eq:bbns}  
\end{align}
where $\mathcal{O}_i $ are the operators from the effective Hamiltonian that describe the underlying weak decay. The $ F_j^{B\rightarrow D}$ form factors and the pion light-cone distribution amplitude (LCDA) $\Phi_{\pi}(u)$, with momentum fractions $u$ and $1-u$ shared among the pion constituents, are the non-perturbative inputs. The hard-scattering kernels $T_{ij}(u)$, on the other hand, can be evaluated in a perturbative expansion in the strong coupling, and are known in QCD to NLO accuracy~\cite{Beneke:2000ry}. Yet it is interesting to go beyond NLO in $B\rightarrow D\pi$ transitions: Since the contribution at NLO is colour suppressed and appears alongside small Wilson coefficients, the NNLO corrections may be significant in size. Moreover, since there is neither a colour-suppressed tree amplitude nor penguin contributions, and spectator scattering and weak annihilation are power-suppressed~\cite{Beneke:2000ry}, we have only the vertex kernels to the colour-allowed tree amplitude. A precise theory prediction of this single contribution, together with comparison to experimental data, might give a reliable estimate of the size of power corrections in the QCDF framework.

The evaluation of Feynman diagrams that contribute to the NNLO hard-scattering kernel amounts to the computation of $\sim 70$ two-loop diagrams. By using contemporary techniques to evaluate multi-loop integrals,
the two-loop Feynman diagrams are reduced to a small set of a few dozens of master integrals.
A powerful method to evaluate the latter analytically are differential equations~\cite{Kotikov:1990kg,Kotikov:1991pm,Remiddi:1997ny}. This method was recently refined by Henn~\cite{Henn:2013pwa}. Considering that the basis of master integrals is not unique, Henn discovered that in a suitably chosen basis -- denoted as \emph{canonical basis} -- the differential equations can be cast into a form that factorises the dependence on the kinematic variables from that on the number of space-time dimensions~$D$. In this case, the solution is expressed in terms of iterated integrals. This method was recently applied to a number of problems for loop~\cite{Henn:2013tua,Henn:2013woa,Henn:2013nsa,Argeri:2014qva,Henn:2014lfa,Gehrmann:2014bfa,Caola:2014lpa,DiVita:2014pza,vonManteuffel:2014mva,Bell:2014zya} and phase-space~\cite{Hoschele:2014qsa,Zhu:2014fma} integrals. 

To the present day, the construction of the canonical basis is mostly based on experience or experimentation, rather than on a systematic procedure, although developments in this direction have recently become available~\cite{Argeri:2014qva,Lee:2014ioa,Henn:2014qga}. In the future it would be most desirable to have a general algorithm for finding a canonical basis for arbitrary external kinematics and numbers of loops, legs, scales, and space-time dimensions. Therefore, every non-trivial example of a canonical basis is most valuable, and our results contribute towards finding a general algorithm for constructing the canonical basis.

Last but not least, if the master integrals that enter the hard-scattering kernels $T_{ij}(u)$ are written in terms of iterated integrals, the convolution with the pion LCDA in~(\ref{eq:bbns}) simplifies to a large extent. Our results therefore catalyse the steps necessary to obtain the decay amplitudes considerably, and constitute an important step towards the phenomenology of $B\rightarrow D\pi$ decays at NNLO in QCDF.

This paper is organized as follows. In section~\ref{sec:kin} we introduce the kinematics of the two-body decay and present the generic form of the differential equations with respect to the kinematic variables. We proceed by defining Goncharov polylogarithm in section~\ref{sec:ints}, which are a class of iterated integrals suited to describe the solutions to the differential equations. In section~\ref{sec:basis} the canonical basis is defined and the expressions for the master integrals in this basis are presented. We also elaborate on strategies to find a canonical basis. The boundary conditions for the integrals are discussed in section~\ref{sec:boundary} and the results are presented in section~\ref{sec:results}. In section~\ref{sec:checks} we comment on the performed cross-checks 
before concluding in section~\ref{sec:conclusion}. In appendix~\ref{app:Atilde} we collect the matrices that contain all relevant information on the differential equations. The analytic results of all master integrals are also available electronically~\cite{electronic}.


\section{Kinematics}
\label{sec:kin}

We consider the kinematics of the decay $\bar B^0 \rightarrow D^{+}\pi^{-}$, which emerges from the underlying weak transition $b\rightarrow c \bar u d$.
 A sample of Feynman diagrams contributing to the two-loop hard-scattering kernels is given in figure~\ref{fig:sample}. The complete set of diagrams consists of those shown in figures~15 and~16 of~\cite{Beneke:2000ry}, supplemented by gluon self-energy insertions in one-loop diagrams.
 All external momenta are taken to be incoming throughout this work. $q_4$ and $q_3$ denote the external momenta of the $b$ and the $c$ quark, respectively, which fulfill the on-shell constraints $q_{4,3}^2=m_{b,c}^2$.
 The constituents of the pion share the momentum $q$ with $q_1= u q$ and $q_2 =(1-u) q\equiv \bar u q$, where $u\in [0,1]$ is the momentum fraction of the quarks inside the pion entering eq.~\eqref{eq:bbns} in a convolution of the hard-scattering kernel with the pion LCDA. We consider the pion to be massless, i.e.\ $q^2= q_{1,2}^2= 0$. 
 Due to the linear dependence of the momenta, $q_1 +q_2 = q = -q_3 -q_4$, the kinematics is completely determined by two of the on-shell conditions and one additional kinematic invariant, for instance
\begin{align}
 q_{4}^2=m_{b}^2 \, , \quad q_{3}^2=m_{c}^2\, ,\quad q_3 q_4 = - \frac{1}{2} (m_b^2+m_c^2) \,.  \label{eq:invar}
\end{align}
We apply commonly used multi-loop techniques which include integration-by-parts identities~\cite{Tkachov:1981wb,Chetyrkin:1981qh} and the Laporta algorithm~\cite{Laporta:2001dd}, and reduce the two-loop Feynman diagrams to master integrals~\cite{Anastasiou:2004vj,Smirnov:2008iw}. Furthermore, we construct the differential equation of the latter with respect to kinematic variables. In the derivation of eq.~\eqref{eq:bbns} the charm quark was assumed to be heavy. Hence, the ratio $m_c/m_b$ remains fixed in the heavy-quark limit and our master integrals depend on two scales:
the momentum fraction $u$ and the ratio of the heavy quark masses $z\equiv m_c^2/m_b^2$.
They are further functions of the kinematic invariants~\eqref{eq:invar}
\begin{align}
 C(u,z)=  C(u,q_3^2(z),(q_4 q_3)(z),q_4^2(z),z) \, .
\end{align}
Thus, the total derivative of a generic master integral $C$ with respect to $u$ is given by 
\begin{align}
 \frac{ dC}{du} =   \frac{\partial C}{ \partial u} \, ,
\end{align}
whereas the one in $z$ reads
\begin{align}
 \frac{d C}{d z}= \frac{\partial C}{\partial z}+ \frac{\partial C}{\partial q_3^2 } \frac{ d q_3^2 }{d z}+ \frac{\partial C}{\partial (q_3 q_4)}\frac{ d (q_3 q_4) }{d z}+ \frac{\partial C}{\partial q_4^2}\frac{ d q_4^2 }{d z}\, . \label{eq:dglz}
\end{align}
The computation of $\partial C / \partial z$ is straightforward.
The partial derivatives of $C$ with respect to the kinematics on the r.h.s.\ of eq.~\eqref{eq:dglz} can be expressed in terms of partial derivatives with respect to the momenta $q_{3,\mu}$ and $q_{4,\mu}$~\cite{Argeri:2007up}, which can be easily carried out. Note that the last term on the r.h.s.\ vanishes since $d q_4^2/d z =0$. We finally obtain 
\begin{align}
  \frac{d C}{d z}= \frac{\partial C}{\partial z}-\frac{1}{1-z} \left(q_{3,\mu} \frac{\partial C}{\partial q_{3,\mu}}+ q_{4,\mu} \frac{\partial C}{\partial q_{3,\mu}} \right) \, .
\end{align}
This is the differential equation with respect to $z$ valid for a generic master integral $C(u,z)$.
\FIGURE[t!]{
  \begin{minipage}{0.3\textwidth}
 \includegraphics[width=1.0\textwidth]{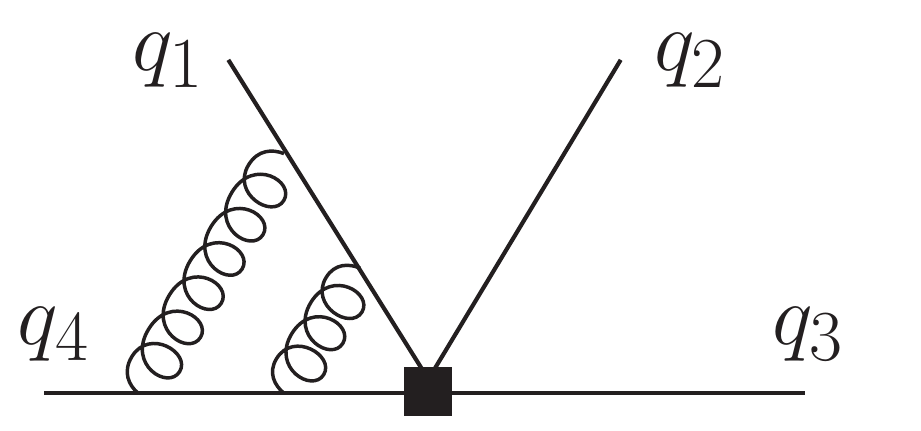} \vspace{0.3mm}
\end{minipage} \hspace{2mm}
  \begin{minipage}{0.3\textwidth} 
\includegraphics[width=1.0\textwidth]{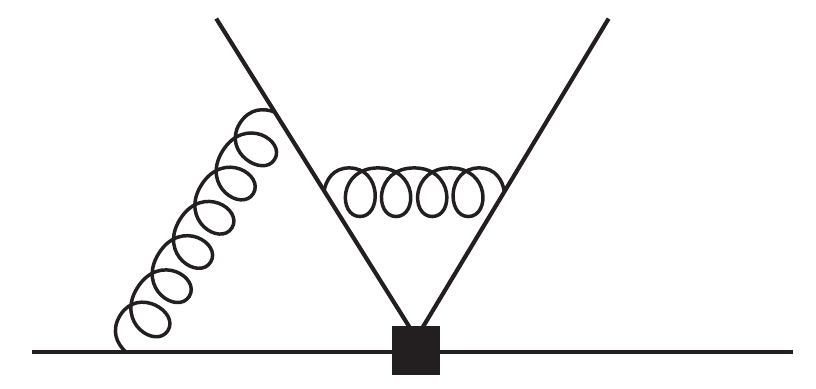} \vspace{0.3mm}
\end{minipage} \hspace{2mm}
  \begin{minipage}{0.3\textwidth}
\includegraphics[width=1.0\textwidth]{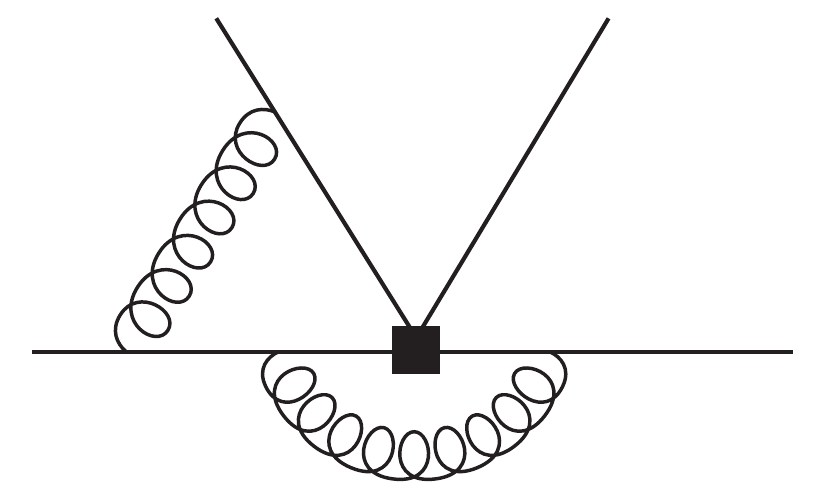}
\end{minipage}
\caption{Sample of Feynman diagrams: $q_4$ and $q_3$ are the momenta of the quark lines with masses $m_b$ and $m_c$, respectively. $q_1=u q$ and $q_2=\bar u q$ are the momenta of the light quark and anti-quark, respectively. $q=q_1+q_2$ is the momentum of the pion. All momenta are incoming. The black square denotes an operator insertion from the weak effective Hamiltonian.
\label{fig:sample}}}


\section{Iterated integrals and Goncharov polylogarithms}
\label{sec:ints}

The classical example of iterated integrals is given by the harmonic polylogarithms (HPLs) \cite{Remiddi:1999ew}. They generalise the ordinary polylogarithms and are defined by
\begin{eqnarray}\label{eq:defHPL}
H_{a_1 , a_2 , \ldots, a_n }(x) = \int_0^x dt \; f_{a_1}(t) \, H_{a_2, \ldots, a_n }(t)  \,, 
\end{eqnarray}
where the parameters $a_i$ can be $0$ or $\pm1$, and $n$ is the weight of the HPL. The integral~(\ref{eq:defHPL}) diverges for HPLs with trailing zeroes. In order to handle HPLs in such cases, one defines $H_{\vec0_n}(x) = \frac{1}{n!} \ln^n (x)$. The weight functions $f_{a_i}(x)$ are simply
\begin{align}\label{eq:weights}
{f_{1}(x) = \frac{1}{1-x}}\,,\qquad  {f_{0}(x) = \frac{1}{x}}\,, \qquad {f_{-1}(x) = \frac{1}{1+x}}\,.
\end{align}
The HPLs fulfil a Hopf algebra according to 
\begin{align}\label{eq:Hopf}
H_{\vec{a}}(x) H_{\vec{b}}(x)=\sum\limits_{\vec c\in \vec{a}\uplus\vec{b}}H_{\vec{c}}(x) \; ,
\end{align}
where $\vec{a}\uplus\vec{b}$ are all possibilities of arranging the elements of $\vec{a}$ and $\vec{b}$ such that the internal order of the elements of $\vec{a}$ and $\vec{b}$ is preserved individually (cf. also~\cite{Maitre:2005uu}). Hence the product of two HPLs of weights $w_1$ and $w_2$ has weight $w_1+w_2$. The Hopf algebra can also be used to extract singular behaviour near $x=0$ or $x=1$. Due to the relation
\begin{align}\label{eq:weightsofnumbers}
H_{0,\ldots,0,1}(1) = \zeta_k
\end{align}
with $k-1$ zeroes and $k>1$, one also assigns the weight $k$ to numbers like $\zeta_k$ and $\pi^k$.

A generalisation of the  HPLs are the Goncharov polylogarithms~\cite{Goncharov:1998kja}, whose definition reads
\begin{align}
G_{a_1 , a_2 , \ldots, a_n }(x) = \int_0^x \frac{dt}{t-a_1} \, G_{a_2, \ldots, a_n }(t) \,  
\label{eq:defGonch}
\end{align}
and $G_{\vec0_n}(x) = H_{\vec0_n}(x)$. They fulfil a Hopf algebra that has the same structure as~(\ref{eq:Hopf}), and allow for more general weights $a_i$ than just $0$ or $\pm 1$. In particular, in multi-scale problems the argument $x$ can be represented by one scale, and the remaining scales are comprised in the weights $a_i$. In our problem at hand, it is most convenient to choose $u$ as the argument of the Goncharov polylogarithm whenever there is a dependence on this scale, bearing in mind that this choice simplifies a subsequent convolution with the light-cone distribution amplitude, which in a Gegenbauer expansion is a $u$-dependent polynomial. In this case the weights are either integer~($0,\pm 1$) or one of the following six $z$-dependent weights\footnote{The analytic results in section~\ref{sec:results} contain only $a_1 - a_4$. The results of the ``mass-flipped'' integrals (see section~\ref{sec:results} and~\cite{electronic}) contain also $a_5$ and $a_6$.},
\begin{align}\label{eq:weightsghoncharov}
a_1 = & \frac{1}{1-z}\, , &  a_3 = & \frac{1}{1-\sqrt{z}}\, , &  a_5 = & \frac{\sqrt{z}}{\sqrt{z}-1} \, , \nnb \\
a_2 = & \frac{z}{z-1}\, , &  a_4 = & \frac{1}{1+\sqrt{z}}\, , &  a_6 = & \frac{\sqrt{z}}{\sqrt{z}+1} \, .
\end{align}
Goncharov polylogarithms that do not depend on $u$ are written in terms of integer weights and argument $z$ or $\sqrt{z}$. Products of Goncharov polylogarithms of the same argument are expanded by means of the Hopf algebra.


\section{The canonical basis}
\label{sec:basis}

We work in dimensional regularisation with $D=4-2\eps$ and evaluate the two-loop, two-scale master integrals by applying the method proposed by Henn~\cite{Henn:2013pwa}. Considering a specific power in the $\eps$-expansion of a master integral, the associated function is called uniform if each summand has the same weight. Moreover, a uniform function is called pure, if its derivative with respect to any one of its arguments yields a uniform function whose weight is lowered by one unit. 

The proposal in~\cite{Henn:2013pwa} now states that a basis $\vec C$ of master integrals can be found such that the system of differential equations in the kinematic variables $x_j$ is given by
\begin{align}
 d_i \vec C(x_j ,\eps) = \eps \,A_i(x_j) \vec C(x_j,\eps) \, , \label{eq:dglcanonical}
\end{align}
where $d_i \equiv d/dx_i$. The $\vec C(x_j,\eps)$ denote the $N$ master integrals and $A_i(x_j)$ are $N\times N$ matrices which are independent of $\eps$.
It turns out that eq.~\eqref{eq:dglcanonical} can be expressed in a compact form
\begin{align}
 d \vec C(x_j,\eps) = \eps \, \big( d \, \tilde{A}(x_j) \big) \vec C(x_j,\eps)  \, , \label{eq:dglcanonical2}
\end{align}
with the function $\tilde{A}$ determined by the differential $d_i \tilde{A} = A_i$. We note that $\tilde{A}$, together with the boundary conditions, completely determines the solution to a master integral. The master integrals in such a basis have in turn several pleasant features: First, the solution decouples order-by-order in the $\eps$-expansion. Second, it is given by pure functions to all orders in $\eps$. Consequently, assigning a weight $-1$ to each power of the expansion parameter $\eps$ and multiplying each master integral by an appropriate power of $\eps$ renders the total weight of the master integral to be zero to all orders. Third, the solution can be expressed in terms of iterated integrals. If the coefficients $A_i(x_j)$ are rational functions of the $x_j$, the Goncharov polylogarithms discussed above represent a suitable class of iterated integrals to describe the master integrals. We will refer to such a basis as a {\emph{canonical basis}}.

In the absence of a completely general algorithm for the systematic construction of the canonical basis, the procedure of finding such a basis requires a certain amount of experience and experimentation. In our case, we start from a ``traditional'' basis that consists of undotted and singly-dotted integrals, and compute them up to terms that involve functions of weight two. For this task, alternative approaches like Feynman parameters or Mellin-Barnes representations~\cite{Smirnov:1999gc,Tausk:1999vh} have to be used. Afterwards one plugs these expressions into seemingly more complicated integrals like the ones in figures~\ref{fig:integrals1} and~\ref{fig:integrals2} and investigates if the resulting expressions are uniform or even pure. This method is mostly based on trial and error, but has proven to be successful as we show below.
\FIGURE[p!]{
 \includegraphics[width=0.95  \textwidth]{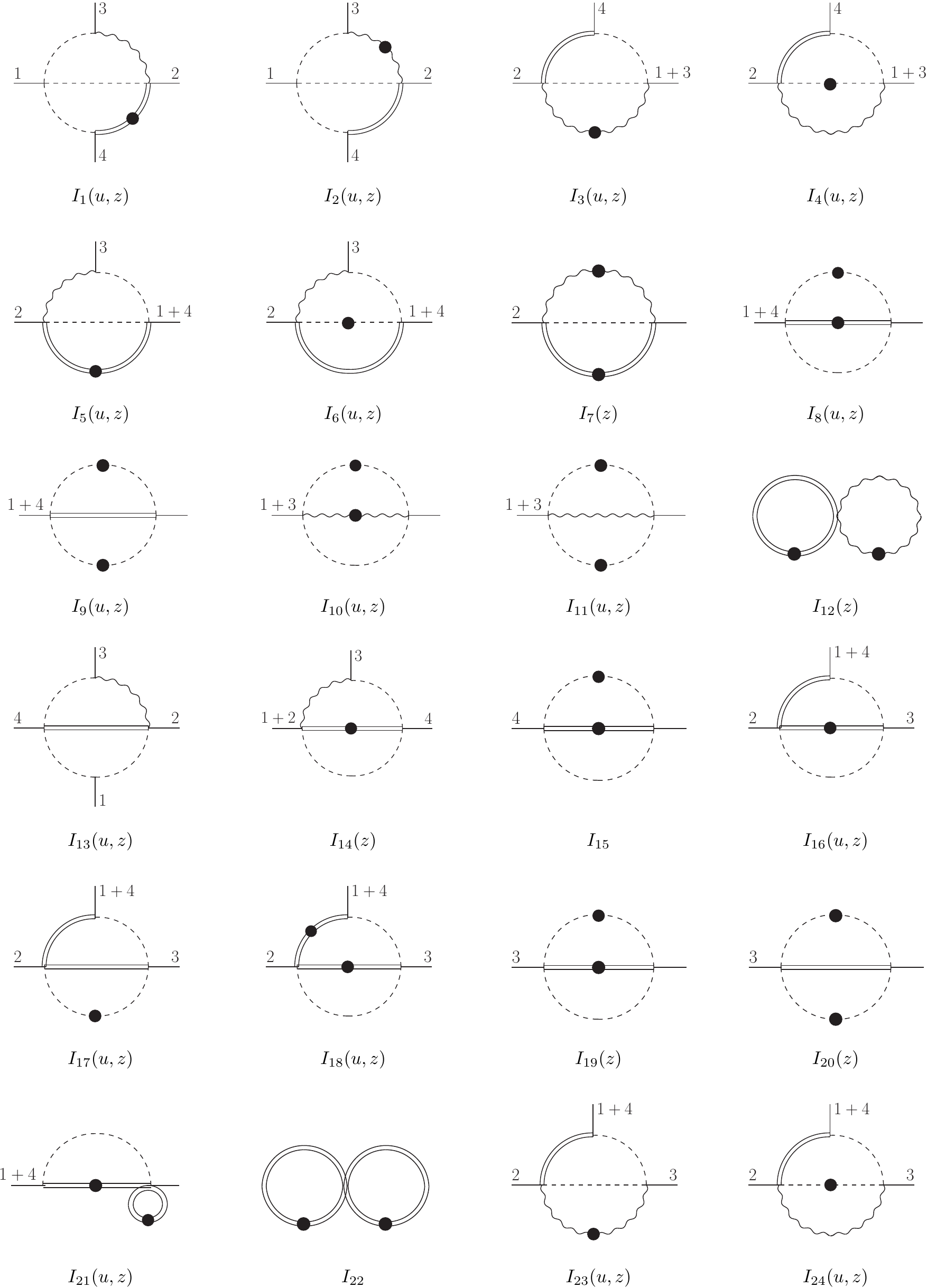}
\caption{Part I of the basic integrals needed in the construction of the canonical basis: $1,\dots ,4$ denote the incoming momenta $q_1,\dots ,q_4$. 
The double/curly/dashed line represents a propagator with mass $m_b/m_c/0$. The dot on a line indicates a squared propagator. 
\label{fig:integrals1}} 
 }
\FIGURE[t!]{
 \includegraphics[width=0.95 \textwidth]{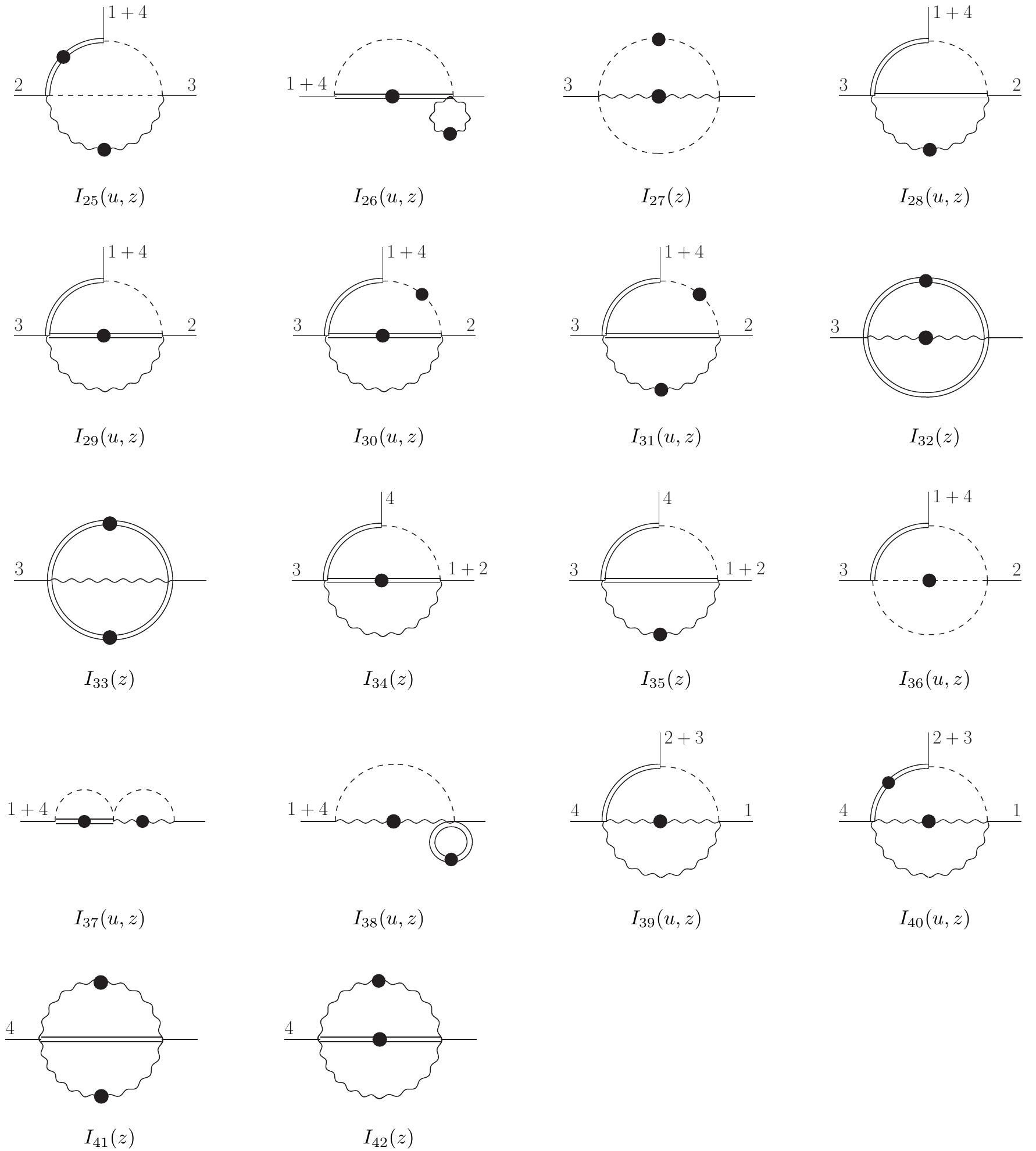}
\caption{Part II of the basic integrals needed in the construction of the canonical basis. All symbols have the same meaning as in figure~\ref{fig:integrals1}.
\label{fig:integrals2}} 
}

In the case at hand, many master integrals can be adopted from several $B\rightarrow \pi\pi$ calculations~\cite{Bell:2009nk,Bell:2007tv,Beneke:2009ek,Huber:2009se}. In order to describe the yet unknown ones in the canonical basis, a set of 39 integrals is needed. We obtain the
following expressions for the canonical master integrals $C_{1-39}$ in terms of the integrals $I_{1-42}$, 
which are defined in figures~\ref{fig:integrals1} and~\ref{fig:integrals2}~($\bar x = 1 - x$).
\allowdisplaybreaks{
\begin{align}
  C_1(u,z) &= \eps^3 \,u  \bar z  \, I_1(u,z)  \, , \\[2mm]
  C_2(u,z)& = \eps^3 \, u (z-1) z  \, I_2(u,z)  \, ,\\[2mm] 
  C_3(u,z) &= \eps^3 \, \bar u  \bar z  \, I_3(u,z)  \,, \\[2mm]
  C_4(u,z) &= \eps^3 \, \bar u   \bar z  \, I_4(u,z)  \, ,\\[2mm] 
  C_5(u,z) &= \eps^3 \, \bar u   (z-1)  \, I_5(u,z)  \, ,\\[2mm]
  C_6(u,z) &= \eps^3 \, \bar u   (z-1)  \, I_6(u,z)  \, ,\\[2mm]
  C_7(z) &= \eps \, (1-\eps)    \bar z  \, I_7(z)  \, ,\\[2mm]
  C_8(u,z) &= \eps^2 \, (\bar u + uz)   \, I_8(u,z)  \, ,\\[1mm]
  C_9(u,z) &= \eps^2 \, u   \bar z   \Big( I_9(u,z) +2 I_8(u,z) \Big)   \, ,\\[1mm]
  C_{10}(u,z) &= \eps^2 \, ( u + \bar u z)  \, I_{10}(u,z)  \, ,\\[1mm]  
  C_{11}(u,z) &= \eps^2 \,  u   (z-1)  \Big( I_{11}(u,z) +2 I_{10}(u,z) \Big)  \, ,\\[1mm]
  C_{12}(z) &= \eps^2 \, I_{12}(z)  \, ,\\[2mm]
  C_{13}(u,z) &= \eps^4 \,  u   \bar z  \, I_{13}(u,z)  \, ,\\[2mm]
  C_{14}(z) &= \eps^3 \,   \bar z  \, I_{14}(z)  \, ,\\[2mm]
  C_{15} &= \eps^2 \, I_{15}  \, ,\\[2mm]
  C_{16}(u,z) &= \eps^3 \, \bar u  \bar z  \, I_{16}(u,z)  \, ,\\[2mm]
  C_{17}(u,z) &= \eps^3 \, \bar u  \bar z  \, I_{17}(u,z)  \, ,\\
  C_{18}(u,z) &= \eps^2 \, (1- \bar u \bar z) \left(  I_{18}(u,z) +\frac{\eps}{m_b^2}I_{17}(u,z)  +\frac{2\eps}{m_b^2}I_{16}(u,z) \right) \, ,\\
  C_{19}(z) &= \eps^2 \,   z  \, I_{19}(z)  \, ,\\[2mm]  
  C_{20}(z) &= \eps^2 \,  \bar z \, \Big( I_{20}(z) +2 I_{19}(z) \Big)  \, ,\\[2mm]
 C_{21}(u,z) &= \eps^2 \,  (1-u \bar z)  \, I_{21}(u,z)  \, ,\\[2mm] 
 C_{22} &= \eps^2   \, I_{22}  \, ,\\[2mm] 
  C_{23}(u,z) &= \eps^3 \, \bar u   \bar z  \, I_{23}(u,z)  \, ,\\[2mm]  
  C_{24}(u,z) &= \eps^3 \, \bar u   \bar z  \, I_{24}(u,z)  \, ,\\
  C_{25}(u,z) &= \eps^2 \, (1- \bar u \bar z) \left(  I_{25}(u,z) +\frac{\eps}{m_b^2}I_{24}(u,z)  +\frac{2\eps}{m_b^2}I_{23}(u,z) \right) \, ,\\
  C_{26}(u,z) &= \eps^2 \,(1-u \bar z)  \, I_{26}(u,z)  \, ,\\[2mm]  
  C_{27}(z) &= \eps^2 \, z \, I_{27}(z)  \, ,\\[2mm]  
  C_{28}(u,z) &= \eps^3 \, \bar u   \bar z  \, I_{28}(u,z)  \, ,\\[2mm] 
  C_{29}(u,z) &= \eps^3 \, \bar u \bar z   \, I_{29}(u,z)  \, ,\\ 
 C_{30}(u,z) &=\frac{1}{2} \eps^2 \, u\bar u   \bar z^2  \, \left(  I_{31}(u,z)+I_{30}(u,z) -\frac{1-\eps}{\eps}\frac{1}{ m_b^2 u \bar z} I_{7}(z) \right) \, ,\\
 C_{31}(z) &= \eps^2 \, z  \, I_{32}(z)  \, ,\\
 C_{32}(z) &= \eps^2 \, \sqrt{z}  \, \Big( I_{33}(z) + 2 I_{32}(z)\Big) \, ,\\ 
 C_{33}(z) &= \eps^3 \,\bar z \, I_{34}(z)  \, ,\\[2mm] 
C_{34}(z) &= \eps^3 \,\bar  z \, I_{35}(z)  \, ,\\[2mm] 
 C_{35}(u,z) &= \eps^3 \, \bar u \bar z \, I_{36}(u,z)  \, ,\\[2mm] 
 C_{36}(u,z) &= \eps^2 \, (1-u \bar z)^2 \, I_{37}(u,z)  \, ,\\[2mm] 
 C_{37}(u,z) &= \eps^2 \, (1-u \bar z) \, I_{38}(u,z)  \, ,\\[2mm] 
 C_{38}(u,z) &= \eps^3 \,  u \bar z  \, I_{39}(u,z)  \, ,\label{eq:C38}\\
 C_{39}(u,z) &= \eps^2  \,    \Big\{ u \bar z \,\left[1-(1-u \bar z )p\right] I_{40}(u,z)  \nonumber \\ &\quad\quad -\frac{1}{m_b^2} \left( \sqrt{z} -\frac{1-(1-u \bar z) p}{2} \right) \Big(I_{41}(z)+2 I_{42}(z) \Big)  \Big\}\, , \label{eq:C39}
\end{align}}
with
\begin{align}
p= \frac{1-\sqrt{(2-u\bar z)^2-4\bar z(1-u\bar z)}}{1-u\bar z} \; .
\end{align}
Note that  the master integrals have to be evaluated to $\mathcal{O}(\eps^4)$ since the two-loop amplitude contains poles up to $1/\eps^4$ stemming from the infrared and ultraviolet regions. A few exceptions are $C_{26,38}$ and $C_{39}$ which only enter the hard-scattering kernel to order $\mathcal{O}(\eps^3)$ and $\mathcal{O}(\eps^2)$, respectively.


\section{Boundary conditions}
\label{sec:boundary}

Before we present the differential equations, we specify the boundary conditions that are used to completely fix the solution. In the simplest cases, the master integrals vanish in a specific kinematic point. This is the case for $C_{13,38,39}$, which vanish in $u=0$, whereas $C_{3,4,5,6,16,17,23,24,28,29,30,35}$ vanish in $u=1$. Moreover, $C_{19,31,32}$ vanish in $z=0$, whereas $C_{7,14,33,34}$ vanish in $z=1$. In other cases we find special relations between integrals, that hold either in general, or in certain kinematic points, and can be used as boundary conditions. Examples are the relation $C_{26} = z^{-\ep} \, C_{21}$, or the following relations that hold in $u=1$,
\begin{align}
C_{8} & \stackrel{u\to 1}{\longrightarrow} C_{19} \, , & C_{10} & \stackrel{u\to 1}{\longrightarrow} C_{19}^{\leftrightarrow} \, , \nnb \\
C_{9} & \stackrel{u\to 1}{\longrightarrow} C_{20} \, , & C_{11} & \stackrel{u\to 1}{\longrightarrow} C_{20}^{\leftrightarrow} \, , \label{eq:bc}
\end{align}
where the symbol ``$\leftrightarrow$'' is used for the corresponding ``mass-flipped'' integral, in which $m_c \leftrightarrow m_b$ and $q_3 \leftrightarrow q_4$, see section~\ref{sec:results} for more details. Hence, the integrals $C_{19,20}^{\leftrightarrow}$ can be easily obtained from $C_{19,20}$ or from~\cite{electronic}. Relations that have a similar structure than~(\ref{eq:bc}) hold in $z=1$ for
\begin{align}
C_{12} & \stackrel{z\to 1}{\longrightarrow} C_{22} \, , & C_{27} & \stackrel{z\to 1}{\longrightarrow} C_{15} \, .
\end{align}
For the remaining integrals we either use that they assume simple, closed forms that are valid to all orders in the $\eps$-expansion, or asymptotic forms as $u \to 0$ or $z\to 0$. Examples of the former type are (see below in section~\ref{sec:results} for the precise definition of $\tilde C_i$)
\begin{align}
\tilde C_{15} & = - \frac{\Gamma^4(1-\ep)\Gamma(1-4\ep)\Gamma (1+\ep) \Gamma (1+2 \ep)}{4 \Gamma (1-3\ep)\Gamma(1-2\ep)} \, , \nnb \\
\tilde C_{22} & = \Gamma^2(1-\ep)\Gamma^2(1+\ep) \, , \nnb \\
\tilde C_{36} & = \left[ - \frac{\ep \,(1-u \bar z)}{(1-\ep)} \,\Gamma(1-\ep) \, \Gamma(1+\ep)  \; \pFq{2}{1}{1,1+\ep}{2-\ep}{\bar u + uz}\right]  \nnb \\ & \times  \left[ - \frac{\ep \, z^{-\ep} \,(1-u \bar z)}{(1-\ep) z} \,\Gamma(1-\ep) \, \Gamma(1+\ep)  \; \pFq{2}{1}{1,1+\ep}{2-\ep}{u+\frac{\bar u}{z}}\right] \, ,
\end{align}
where for $C_{36}$ we give the result for each loop separately, such that also the boundary conditions for $C_{21,37}$ can be read off. Asymptotic expansions as $u \to 0$ or $z\to 0$ were derived by means of {\tt MBasymptotics.m}~\cite{MBasymptotics} for
\allowdisplaybreaks{
\begin{align}
\tilde C_{20} \stackrel{z\to 0}{=}& -1-\frac{2 \pi ^2 }{3} \, \epcolor{\ep^2} + 2 \zeta_3 \, \epcolor{\ep^3} -\frac{5 \pi ^4}{18} \, \epcolor{\ep^4} + {\cal O}(\ep^5,z) \, , \nnb \\
\tilde C_{1} \stackrel{u\to 0}{=}& \frac{1}{24}+\epcolor{\ep}\,[-\frac{1}{6} \, \ln(u)+\frac{1}{8} \, G_{0}(z)-\frac{1}{6} \, G_{1}(z)+\frac{1}{4} \, i\pi] \nnb \\
+& \epcolor{\ep^2}\,[\frac{1}{3} \, \ln^2(u)+(\frac{2}{3} \, G_{1}(z)-\frac{1}{2} \, G_{0}(z)- \, i\pi)\ln(u)
+\frac{3}{4} \, i\pi \, G_{0}(z)- \, i\pi \, G_{1}(z)+\frac{3}{8} \, G_{0,0}(z) \nnb \\
-& \frac{1}{2} \, G_{0,1}(z)-\frac{1}{2} \, G_{1,0}(z)+\frac{2}{3} \, G_{1,1}(z)-\frac{37\pi^2}{72}] \nnb \\
+& \epcolor{\ep^3}\,[-\frac{4}{9}\ln^3(u)+( G_{0}(z)-\frac{4}{3} \, G_{1}(z)+2 \, i\pi)\ln^2(u)
+(4 \, i\pi \, G_{1}(z)-3 \, i\pi \, G_{0}(z) \nnb \\
-& \frac{3}{2} \, G_{0,0}(z)+2 \, G_{0,1}(z)+2 \, G_{1,0}(z)-\frac{8}{3} \, G_{1,1}(z)+\frac{37\pi^2}{18})\ln(u)-\frac{37\pi^2}{24} \, G_{0}(z)\nnb \\
+& \frac{37\pi^2}{18} \, G_{1}(z)+\frac{5}{4} \, i\pi \, G_{0,0}(z)-3 \, i\pi \, G_{0,1}(z)-2 \, i\pi \, G_{1,0}(z)
+4 \, i\pi \, G_{1,1}(z)+\frac{1}{8} \, G_{0,0,0}(z) \nnb \\
-& \frac{3}{2} \, G_{0,0,1}(z)-\frac{3}{2} \, G_{0,1,0}(z)+2 \, G_{0,1,1}(z)-\frac{1}{2} \, G_{1,0,0}(z)+2 \, G_{1,0,1}(z)+2 \, G_{1,1,0}(z)\nnb \\
-& \frac{8}{3} \, G_{1,1,1}(z)-\frac{17}{6} \, \zeta_3-\frac{7}{12} \, i\pi^3] \nnb \\
+& \epcolor{\ep^4}\,[\frac{4}{9} \, \ln^4(u)+(\frac{16}{9} \, G_{1}(z)-\frac{4}{3} \, G_{0}(z)-\frac{8}{3}\, i\pi)\ln^3(u)
+(6 \, i\pi \, G_{0}(z)-8 \, i\pi \, G_{1}(z) \nnb \\
+& 3 \, G_{0,0}(z)-4 \, G_{0,1}(z)-4 \, G_{1,0}(z)+\frac{16}{3} \, G_{1,1}(z)-\frac{37\pi^2}{9})\ln^2(u)+(\frac{37\pi^2}{6} \, G_{0}(z)\nnb \\
-& \frac{74\pi^2}{9} \, G_{1}(z)-5 \, i\pi \, G_{0,0}(z)+12 \, i\pi \, G_{0,1}(z)+8 \, i\pi \, G_{1,0}(z)-16 \, i\pi \, G_{1,1}(z)+\frac{34}{3} \, \zeta_3 \nnb \\
-&\frac{1}{2} \, G_{0,0,0}(z)+6 \, G_{0,0,1}(z)+6 \, G_{0,1,0}(z)-8 \, G_{0,1,1}(z)+2 \, G_{1,0,0}(z)-8 \, G_{1,0,1}(z)\nnb \\
-& 8 \, G_{1,1,0}(z)+\frac{32}{3} \, G_{1,1,1}(z)+\frac{7}{3} \, i\pi^3)\ln(u)-\frac{17}{12} \, i\pi^3 \, G_{0}(z)+2 \, i\pi^3 \, G_{1}(z)-8 \, G_{1,0,1,1}(z)\nnb \\
-& \frac{35\pi^2}{24} \, G_{0,0}(z)+\frac{37\pi^2}{6} \, G_{0,1}(z)+3\pi^2 \, G_{1,0}(z)-\frac{74\pi^2}{9} \, G_{1,1}(z)+\frac{3}{4} \, i\pi \, G_{0,0,0}(z)\nnb \\
-& 5 \, i\pi \, G_{0,0,1}(z)- 4 \, i\pi \, G_{0,1,0}(z)+12 \, i\pi \, G_{0,1,1}(z)-2 \, i\pi \, G_{1,0,0}(z)+8 \, i\pi \, G_{1,0,1}(z)\nnb \\
+& 6 \, i\pi \, G_{1,1,0}(z)-16 \, i\pi \, G_{1,1,1}(z)+\frac{3}{8} \, G_{0,0,0,0}(z)-\frac{1}{2} \, G_{0,0,0,1}(z)-\frac{3}{2} \, G_{0,0,1,0}(z)\nnb \\
+& 6 \, G_{0,0,1,1}(z)+\frac{1}{2} \, G_{0,1,0,0}(z)+6 \, G_{0,1,0,1}(z)+6 \, G_{0,1,1,0}(z)-8 \, G_{0,1,1,1}(z)+\frac{82\pi^4}{135}\nnb \\
-& \frac{1}{2} \, G_{1,0,0,0}(z)+2 \, G_{1,0,0,1}(z)+3 \, G_{1,0,1,0}(z)-8 \, G_{1,1,0,1}(z)-8 \, G_{1,1,1,0}(z)\nnb \\
+& \frac{32}{3} \, G_{1,1,1,1}(z)-\frac{17}{2} \, G_{0}(z)\zeta_3+\frac{34}{3} \, G_{1}(z) \zeta_3
-12 \, i\pi\zeta_3]
+ {\cal O}(\ep^5,u) \, , \\[1.5em]
\tilde C_{2} \stackrel{u\to 0}{=}& \frac{1}{24} + \epcolor{\ep}\,[-\frac{1}{6} \, \ln(u)-\frac{1}{24} \, G_{0}(z)-\frac{1}{6} \, G_{1}(z)-\frac{1}{12} \, i\pi]\nnb \\
+& \epcolor{\ep^2}\,[\frac{1}{3} \, \ln^2(u)+(\frac{1}{6} \, G_{0}(z)+\frac{2}{3} \, G_{1}(z)+\frac{1}{3} \, i\pi)\ln(u)
+\frac{1}{12} \, i\pi \, G_{0}(z)+\frac{1}{3} \, i\pi \, G_{1}(z)\nnb \\
+& \frac{1}{24} \, G_{0,0}(z)+\frac{1}{6} \, G_{0,1}(z)+\frac{1}{6} \, G_{1,0}(z)+\frac{2}{3} \, G_{1,1}(z)+\frac{11\pi^2}{72}]\nnb \\
+& \epcolor{\ep^3}\,[-\frac{4}{9}\ln^3(u)+(-\frac{1}{3} \, G_{0}(z)-\frac{4}{3} \, G_{1}(z)
-\frac{2}{3} \, i\pi)\ln^2(u)+(-\frac{1}{3} \, i\pi \, G_{0}(z)-\frac{4}{3} \, i\pi \, G_{1}(z)\nnb \\
-& \frac{1}{6} \, G_{0,0}(z)-\frac{2}{3} \, G_{0,1}(z)-\frac{2}{3} \, G_{1,0}(z)-\frac{8}{3} \, G_{1,1}(z)-\frac{11\pi^2}{18})\ln(u)
-\frac{11\pi^2}{72} \, G_{0}(z)\nnb \\
-& \frac{11\pi^2}{18} \, G_{1}(z)-\frac{1}{12} \, i\pi \, G_{0,0}(z)-\frac{1}{3} \, i\pi \, G_{0,1}(z)+\frac{2}{3} \, i\pi \, G_{1,0}(z)-\frac{4}{3} \, i\pi \, G_{1,1}(z)\nnb \\
-& \frac{1}{24} \, G_{0,0,0}(z)-\frac{1}{6} \, G_{0,0,1}(z)-\frac{1}{6} \, G_{0,1,0}(z)
-\frac{2}{3} \, G_{0,1,1}(z)+\frac{5}{6} \, G_{1,0,0}(z)-\frac{2}{3} \, G_{1,0,1}(z)\nnb \\
-& \frac{2}{3} \, G_{1,1,0}(z)-\frac{8}{3} \, G_{1,1,1}(z)-\frac{17}{6} \, \zeta_3-\frac{1}{4} \, i\pi^3]\nnb \\
+ & \epcolor{\ep^4}\,[\frac{4}{9} \, \ln^4(u)+(\frac{4}{9} \, G_{0}(z)+\frac{16}{9} \, G_{1}(z)+\frac{8}{9} \, i\pi)\ln^3(u)
+(\frac{2}{3} \, i\pi \, G_{0}(z)+\frac{8}{3} \, i\pi \, G_{1}(z)\nnb \\
+&\frac{1}{3} \, G_{0,0}(z)+\frac{4}{3} \, G_{0,1}(z)+\frac{4}{3} \, G_{1,0}(z)+\frac{16}{3} \, G_{1,1}(z)+\frac{11\pi^2}{9})\ln^2(u)
+(\frac{11\pi^2}{18} \, G_{0}(z)\nnb \\
+&\frac{22\pi^2}{9} \, G_{1}(z)+\frac{1}{3} \, i\pi \, G_{0,0}(z)+\frac{4}{3} \, i\pi \, G_{0,1}(z)-\frac{8}{3} \, i\pi \, G_{1,0}(z)+\frac{16}{3} \, i\pi \, G_{1,1}(z)\nnb \\
+& \frac{1}{6} \, G_{0,0,0}(z)+\frac{2}{3} \, G_{0,0,1}(z)+\frac{2}{3} \, G_{0,1,0}(z)
+\frac{8}{3} \, G_{0,1,1}(z)-\frac{10}{3} \, G_{1,0,0}(z)+\frac{8}{3} \, G_{1,0,1}(z)\nnb \\
+& \frac{8}{3} \, G_{1,1,0}(z)+\frac{32}{3} \, G_{1,1,1}(z)+\frac{34}{3} \, \zeta_3+ i\pi^3)\ln(u)
+\frac{1}{4} \, i\pi^3 \, G_{0}(z)+\frac{2}{3} \, i\pi^3 \, G_{1}(z)\nnb \\
+&\frac{11\pi^2}{72} \, G_{0,0}(z)+\frac{11\pi^2}{18} \, G_{0,1}(z)-\frac{5\pi^2}{9} \, G_{1,0}(z)+\frac{22\pi^2}{9} \, G_{1,1}(z)
+\frac{1}{12} \, i\pi \, G_{0,0,0}(z)\nnb \\
+&\frac{1}{3} \, i\pi \, G_{0,0,1}(z)-\frac{2}{3} \, i\pi \, G_{0,1,0}(z)
+\frac{4}{3} \, i\pi \, G_{0,1,1}(z)+\frac{10}{3} \, i\pi \, G_{1,0,0}(z)-\frac{8}{3} \, i\pi \, G_{1,0,1}(z)\nnb \\
-& \frac{14}{3} \, i\pi \, G_{1,1,0}(z)+\frac{16}{3} \, i\pi \, G_{1,1,1}(z)+\frac{1}{24} \, G_{0,0,0,0}(z)
+\frac{1}{6} \, G_{0,0,0,1}(z)+\frac{1}{6} \, G_{0,0,1,0}(z)\nnb \\
+&\frac{2}{3} \, G_{0,0,1,1}(z)-\frac{5}{6} \, G_{0,1,0,0}(z)+\frac{2}{3} \, G_{0,1,0,1}(z)+\frac{2}{3} \, G_{0,1,1,0}(z)
+\frac{8}{3} \, G_{0,1,1,1}(z)-\frac{4}{3} \, i\pi \, \zeta_3\nnb \\
+& \frac{19}{6} \, G_{1,0,0,0}(z)-\frac{10}{3} \, G_{1,0,0,1}(z)-\frac{7}{3} \, G_{1,0,1,0}(z)+\frac{8}{3} \, G_{1,0,1,1}(z)
-\frac{16}{3} \, G_{1,1,0,0}(z)+\frac{49\pi^4}{135}\nnb \\
+& \frac{8}{3} \, G_{1,1,0,1}(z)+\frac{8}{3} \, G_{1,1,1,0}(z)+\frac{32}{3} \, G_{1,1,1,1}(z)
+\frac{17}{6} \, G_{0}(z) \, \zeta_3+\frac{34}{3} \, G_{1}(z) \, \zeta_3]
+ {\cal O}(\ep^5,u) \, , \\[1.5em]
\tilde C_{18} \stackrel{u\to 0}{=}&\epcolor{\ep^2}\,[G_{1}(z) \, \ln(u) - G_{0,1}(z) + G_{1,1}(z)]\nnb \\
+& \epcolor{\ep^3}\,[(G_{0,1}(z)-6 \, G_{1,1}(z)) \ln(u)- G_{1}(z) \ln^2(u)+\frac{\pi^2}{6} \, G_{1}(z)+5 \, G_{0,1,1}(z)-6 \, G_{1,1,1}(z)]\nnb \\
+&\epcolor{\ep^4}\,[\frac{2}{3}\ln^3(u) \, G_{1}(z)+(6 \, G_{1,1}(z)- \, G_{0,1}(z))\ln^2(u)
+(\frac{2\pi^2}{3} \, G_{1}(z)+ G_{0,0,1}(z)\nnb \\
-& 6 \, G_{0,1,1}(z)-4 \, G_{1,0,1}(z)+28 \, G_{1,1,1}(z)) \, \ln(u)+5 \, \zeta_3 \, G_{1}(z)-\frac{5\pi^2}{6} \, G_{0,1}(z)\nnb \\
+& \frac{\pi^2}{3} \, G_{1,1}(z)- \, G_{0,0,0,1}(z)+ \, G_{0,0,1,1}(z)+4 \, G_{0,1,0,1}(z)-22 \, G_{0,1,1,1}(z)
+4 \, G_{1,0,0,1}(z)\nnb \\
-& 4 \, G_{1,0,1,1}(z)+2 \, G_{1,1,0,1}(z)+28 \, G_{1,1,1,1}(z)]
+ {\cal O}(\ep^5,u) \, , \\[1.5em]
\tilde C_{25} \stackrel{u\to 0}{=}&\epcolor{\ep^2}\,[\frac{\ln^2(u)}{2} +(G_{1}(z)-G_{0}(z)) \ln(u)+ \, G_{0,0}(z)- \, G_{0,1}(z)- \, G_{1,0}(z)+ \, G_{1,1}(z)+\frac{\pi^2}{2}]\nnb \\
+& \epcolor{\ep^3}\, [-\ln^3(u)+(\frac{3}{2} \, G_{0}(z)-3 \, G_{1}(z)) \, \ln^2(u)+(3 \, G_{0,1}(z)+2 \, G_{1,0}(z)
-6 \, G_{1,1}(z)\nnb \\
-& \pi^2) \, \ln(u)-\frac{\pi^2}{2} \, G_{0}(z)-\frac{7\pi^2}{6} \, G_{1}(z)-3 \, G_{0,0,0}(z)+ \, G_{0,1,0}(z)
+3 \, G_{0,1,1}(z)\nnb \\
+&2 \, G_{1,0,1}(z)+2 \, G_{1,1,0}(z)-6 \, G_{1,1,1}(z)]\nnb \\
+ &\epcolor{\ep^4}\, [\frac{7}{6} \, \ln^4(u)+(\frac{14}{3} \, G_{1}(z)-\frac{5}{3} \, G_{0}(z)) \, \ln^3(u)+(\frac{1}{2} \, G_{0,0}(z)-5 \, G_{0,1}(z)
-3 \, G_{1,0}(z)\nnb \\
+& 14 \, G_{1,1}(z)+\frac{7\pi^2}{3}) \, \ln^2(u)+(5\pi^2 \, G_{1}(z)-\frac{5\pi^2}{3} \, G_{0}(z)- \, G_{0,0,0}(z)+ \, G_{0,0,1}(z)\nnb \\
-& 10 \, G_{0,1,1}(z)+2 \, G_{1,0,0}(z)-6 \, G_{1,0,1}(z)-4 \, G_{1,1,0}(z)+28 \, G_{1,1,1}(z)+6 \, \zeta_3) \, \ln(u)\nnb \\
-& 6 \, \zeta_3 \, G_{0}(z)+3 \, \zeta_3 \, G_{1}(z)+\frac{19\pi^2}{6} \, G_{0,0}(z)-\frac{3\pi^2}{2} \, G_{0,1}(z)-\frac{5\pi^2}{3} \, G_{1,0}(z)
+\frac{16\pi^2}{3} \, G_{1,1}(z)\nnb \\
+& 10 \, G_{0,0,0,0}(z)- \, G_{0,0,0,1}(z)-3 \, G_{0,0,1,0}(z)+ \, G_{0,0,1,1}(z)
-2 \, G_{0,1,0,0}(z)-2 \, G_{0,1,1,0}(z)\nnb \\
-& 10 \, G_{0,1,1,1}(z)-2 \, G_{1,0,0,0}(z)+2 \, G_{1,0,0,1}(z)+2 \, G_{1,0,1,0}(z)-6 \, G_{1,0,1,1}(z)\nnb \\
+& 2 \, G_{1,1,0,0}(z)-4 \, G_{1,1,0,1}(z)-4 \, G_{1,1,1,0}(z)+28 \, G_{1,1,1,1}(z)+\frac{16\pi^4}{15}]
\! + {\cal O}(\ep^5,u) .
\end{align}
}


\section{Results}
\label{sec:results}

In order to facilitate the presentation of the results we write the master integrals as
\begin{align}
C =& \; - \, S_\Gamma^2 \, \left(m_b^2\right)^{D-n} \, \tilde C \, , \label{eq:mastertilde}
\end{align}
with an integer $n$ that denotes the sum of all propagator powers, such that the integral $\tilde C$ is dimensionless. Our integration measure is $\int d^Dk/(2\pi)^D$ per loop and we use the pre-factor
\begin{align}
S_\Gamma =& \frac{1}{\left(4\pi\right)^{D/2} \, \Gamma(1-\ep)} \; .
\end{align}
Besides the integrals defined in section~\ref{sec:basis}, the QCD amplitude also contains the same set of integrals but with $m_c \leftrightarrow m_b$ and $q_3 \leftrightarrow q_4$. We will refer to these as ``mass-flipped'' integrals and denote them as $C^{\leftrightarrow}$, see section~\ref{sec:boundary}. However, we note here that in order to define $\tilde C^{\leftrightarrow}$ we factor out an appropriate power of $m_b$, rather than $m_c$.

As stated earlier the QCD amplitude requires terms of order ${\cal O}(\ep^4)$ for most of the integrals. However, in order to keep the paper at a reasonable length, we only give terms up to order ${\cal O}(\ep^3)$ explicitly below. If desired, terms of weight four can be derived from the $\tilde A$ and the boundary condition, which we actually give to weight four. Moreover, we refrain from presenting the ``mass-flipped'' integrals explicitly. They can be obtained by letting $z \to 1/z$, keeping in mind that analytic continuation is done via $z \to z-i\eta$, with infinitesimal $\eta>0$. We provide the results to all integrals, including the ``mass-flipped'' ones, to order ${\cal O}(\ep^4)$ in electonic form in~\cite{electronic}.

Last but not least, instead of dealing with one large $39 \times 39$ system of equations, we solve each topology separately and therefore deal with several, smaller matrices $\tilde A_i$ which we collect in appendix~\ref{app:Atilde}. This finally puts us in the position to present the analytic results to the $C_{1-39}$.

\subsection{$C_{1}$ -- $C_{12}$}
\label{sec:C01}

We start right away with the largest topology, which contains twelve integrals,
\begin{align}
\vec{C} =& \left\{\tilde C_{1},\tilde C_{2},\tilde C_{3},\tilde C_{4},\tilde C_{5},\tilde C_{6},\tilde C_{7},\tilde C_{8},\tilde C_{9},\tilde C_{10},\tilde C_{11},\tilde C_{12}\right\}\,.
\end{align}
The corresponding matrix is $\tilde A_{1-12}$. Taking into account the boundary conditions specified in the previous section, the solution to the twelve integrals reads
\allowdisplaybreaks{
\begin{align}
\tilde C_{1}=& \frac{1}{24}+\epcolor{\ep}\,[-\frac{1}{6}\,G_{0}(u)+\frac{1}{8}\,G_{0}(z)-\frac{1}{6}\,G_{1}(z)+\frac{1}{4}\,i\pi] \nnb \\
+&\epcolor{\ep^2}\,[-\frac{1}{2}\,G_{0}(z)\,G_{0}(u)+\frac{2}{3}\,G_{1}(z)\,G_{0}(u)-i\pi\,G_{0}(u)+\frac{3}{4}\,i\pi\,G_{0}(z)-i\pi\,G_{1}(z)\nnb\\
+&\frac{1}{2}\,G_{0}(z)\,G_{a_2}(u)-\frac{1}{2}\,G_{1}(z)\,G_{a_2}(u)+\frac{1}{2}\,i\pi\,G_{a_2}(u)+\frac{2}{3}\,G_{0,0}(u)
+\frac{3}{8}\,G_{0,0}(z)-\frac{1}{2}\,G_{0,1}(z) \nnb \\
-&\frac{1}{2}\,G_{1,0}(z)+\frac{2}{3}\,G_{1,1}(z)
-\frac{1}{2}\,G_{a_2,0}(u)-\frac{37\pi^2}{72}] \nnb \\
+&\epcolor{\ep^3}\,[-3\,i\pi\,G_{0}(z)\,G_{0}(u)+4\,i\pi\,G_{1}(z)\,G_{0}(u)-\frac{3}{2}\,G_{0,0}(z)\,G_{0}(u)+2\,G_{0,1}(z)\,G_{0}(u) \nnb \\
+& 2\,G_{1,0}(z)\,G_{0}(u)-\frac{8}{3}\,G_{1,1}(z)\,G_{0}(u)+\frac{37\pi^2}{18}\,G_{0}(u)-\frac{37\pi^2}{24}\,G_{0}(z)
+i\pi\,G_{0}(z)\,G_{1}(u) \nnb \\
+&\frac{37\pi^2}{18}\,G_{1}(z)+\frac{1}{2}\,i\pi\,G_{0}(z)\,G_{a_2}(u)
-2\,i\pi\,G_{1}(z)\,G_{a_2}(u)-\frac{13\pi^2}{12}\,G_{a_2}(u)+2\,G_{0}(z)\,G_{0,0}(u) \nnb \\
-&\frac{8}{3}\,G_{1}(z)\,G_{0,0}(u)+4\,i\pi\,G_{0,0}(u)+G_{1}(u)\,G_{0,0}(z)
-G_{a_2}(u)\,G_{0,0}(z)+\frac{5}{4}\,i\pi\,G_{0,0}(z) \nnb \\
-&\frac{1}{2}\,G_{a_2}(u)\,G_{0,1}(z)-3\,i\pi\,G_{0,1}(z)
-2\,G_{0}(z)\,G_{0,a_2}(u)+2\,G_{1}(z)\,G_{0,a_2}(u)-2\,i\pi\,G_{0,a_2}(u) \nnb \\
-&\frac{1}{2}\,G_{a_2}(u)\,G_{1,0}(z)-2\,i\pi\,G_{1,0}(z)
+2\,G_{a_2}(u)\,G_{1,1}(z)+4\,i\pi\,G_{1,1}(z)-G_{1}(z)\,G_{1,a_1}(u) \nnb \\
+& G_{0}(z)\,G_{1,a_2}(u)-G_{1}(z)\,G_{1,a_2}(u)
+i\pi\,G_{1,a_2}(u)-\frac{1}{2}\,G_{0}(z)\,G_{a_2,0}(u)+2\,G_{1}(z)\,G_{a_2,0}(u) \nnb \\
-& 2\,i\pi\,G_{a_2,0}(u)-\frac{1}{2}\,G_{0}(z)\,G_{a_2,a_2}(u)+\frac{1}{2}\,G_{1}(z)\,G_{a_2,a_2}(u)
-\frac{1}{2}\,i\pi\,G_{a_2,a_2}(u)-\frac{8}{3}\,G_{0,0,0}(u) \nnb \\
+&\frac{1}{8}\,G_{0,0,0}(z)-\frac{3}{2}\,G_{0,0,1}(z)
-\frac{3}{2}\,G_{0,1,0}(z)+2\,G_{0,1,1}(z)+2\,G_{0,a_2,0}(u)-\frac{1}{2}\,G_{1,0,0}(z) \nnb \\
+& 2\,G_{1,0,1}(z)+2\,G_{1,1,0}(z)-\frac{8}{3}\,G_{1,1,1}(z)-G_{1,a_1,0}(u)-G_{1,a_2,0}(u)
+2\,G_{a_2,0,0}(u) \nnb \\
+&\frac{1}{2}\,G_{a_2,a_2,0}(u)-\frac{17}{6}\,\zeta_3-\frac{7}{12}\,i\pi^3] + {\cal O}(\ep^4) \, , \\[1.5em]
\tilde C_{2}=& \frac{1}{24}+\epcolor{\ep}\,[-\frac{1}{6}\,G_{0}(u)-\frac{1}{24}\,G_{0}(z)-\frac{1}{6}\,G_{1}(z)-\frac{1}{12}\,i\pi] \nnb \\
+&\epcolor{\ep^2}\,[\frac{1}{6}\,G_{0}(z)\,G_{0}(u)+\frac{2}{3}\,G_{1}(z)\,G_{0}(u)
+\frac{1}{3}\,i\pi\,G_{0}(u)+\frac{1}{12}\,i\pi\,G_{0}(z)+\frac{1}{3}\,i\pi\,G_{1}(z)\nnb \\
-&\frac{1}{2}\,G_{1}(z)\,G_{a_1}(u)+\frac{2}{3}\,G_{0,0}(u)+\frac{1}{24}\,G_{0,0}(z)+\frac{1}{6}\,G_{0,1}(z)+\frac{1}{6}\,G_{1,0}(z) 
+\frac{2}{3}\,G_{1,1}(z)\nnb \\
-&\frac{1}{2}\,G_{a_1,0}(u)+\frac{11\pi^2}{72}] \nnb \\
+&\epcolor{\ep^3}\,[-\frac{1}{3}\,i\pi\,G_{0}(z)\,G_{0}(u)-\frac{4}{3}\,i\pi\,G_{1}(z)\,G_{0}(u)-\frac{1}{6}\,G_{0,0}(z)\,G_{0}(u)
-\frac{2}{3}\,G_{0,1}(z)\,G_{0}(u) \nnb \\
-&\frac{2}{3}\,G_{1,0}(z)\,G_{0}(u)-\frac{8}{3}\,G_{1,1}(z)\,G_{0}(u)-\frac{11\pi^2}{18}\,G_{0}(u)
-\frac{11\pi^2}{72}\,G_{0}(z)+i\pi\,G_{0}(z)\,G_{1}(u) \nnb \\
-&\frac{11\pi^2}{18}\,G_{1}(z)-\frac{\pi^2}{12}\,G_{a_1}(u)-\frac{2}{3}\,G_{0}(z)\,G_{0,0}(u)-\frac{8}{3}\,G_{1}(z)\,G_{0,0}(u)
-\frac{4}{3}\,i\pi\,G_{0,0}(u) \nnb \\
+& G_{1}(u)\,G_{0,0}(z)-\frac{1}{12}\,i\pi\,G_{0,0}(z)-\frac{1}{2}\,G_{a_1}(u)\,G_{0,1}(z)-\frac{1}{3}\,i\pi\,G_{0,1}(z)
+2\,G_{1}(z)\,G_{0,a_1}(u) \nnb \\
-&\frac{1}{2}\,G_{a_1}(u)\,G_{1,0}(z)+\frac{2}{3}\,i\pi\,G_{1,0}(z)+2\,G_{a_1}(u)\,G_{1,1}(z)-\frac{4}{3}\,i\pi\,G_{1,1}(z)
-G_{1}(z)\,G_{1,a_1}(u) \nnb \\
+& G_{0}(z)\,G_{1,a_2}(u)-G_{1}(z)\,G_{1,a_2}(u)+i\pi\,G_{1,a_2}(u)-\frac{1}{2}\,G_{0}(z)\,G_{a_1,0}(u)+2\,G_{1}(z)\,G_{a_1,0}(u) \nnb \\
+&\frac{1}{2}\,G_{1}(z)\,G_{a_1,a_1}(u)-\frac{8}{3}\,G_{0,0,0}(u)-\frac{1}{24}\,G_{0,0,0}(z)-\frac{1}{6}\,G_{0,0,1}(z)
-\frac{1}{6}\,G_{0,1,0}(z) \nnb \\
-&\frac{2}{3}\,G_{0,1,1}(z)+2\,G_{0,a_1,0}(u)+\frac{5}{6}\,G_{1,0,0}(z)-\frac{2}{3}\,G_{1,0,1}(z)-\frac{2}{3}\,G_{1,1,0}(z)
-\frac{8}{3}\,G_{1,1,1}(z) \nnb \\
-& G_{1,a_1,0}(u)-G_{1,a_2,0}(u)+2\,G_{a_1,0,0}(u)+\frac{1}{2}\,G_{a_1,a_1,0}(u)-\frac{17}{6}\,\zeta_3
-\frac{1}{4}\,i\pi^3]+ {\cal O}(\ep^4) \, , \\[1.5em]
\tilde C_{3}=& \epcolor{\ep^3}\,[i\pi\,G_{a_2}(u)\,G_{0}(z)-G_{a_2,0}(u)\,G_{0}(z)+2\,G_{a_2,a_2}(u)\,G_{0}(z)+\frac{\pi^2}{2}\,G_{0}(z)
+\frac{\pi^2}{3}\,G_{a_2}(u) \nnb \\
+& 2\,G_{a_2}(u)\,G_{0,0}(z)-G_{a_2}(u)\,G_{0,1}(z)-G_{a_2}(u)\,G_{1,0}(z)-2\,G_{1}(z)\,G_{a_2,a_2}(u) \nnb \\
+& 2\,i\pi\,G_{a_2,a_2}(u)+G_{0,0,0}(z)-G_{0,1,0}(z)-2\,G_{a_2,a_2,0}(u)+2\,\zeta_3] + {\cal O}(\ep^4) \, , \\[1.5em]
\tilde C_{4}=& \epcolor{\ep^2}\,[G_{0}(u)\,G_{0}(z)-G_{a_2}(u)\,G_{0}(z)-i\pi\,G_{0}(z)
+G_{1}(z)\,G_{a_2}(u)-i\pi\,G_{a_2}(u)-G_{0,0}(z) \nnb \\
+& G_{0,1}(z)+G_{a_2,0}(u)+\frac{\pi^2}{6}]\nnb \\
+&\epcolor{\ep^3}\,[4\,i\pi\,G_{0}(z)\,G_{0}(u)+3\,G_{0,0}(z)\,G_{0}(u)-4\,G_{0,1}(z)\,G_{0}(u)-2\,G_{1,0}(z)\,G_{0}(u)\nnb \\
-&\frac{\pi^2}{3}\,G_{0}(u)+\frac{3\pi^2}{2}\,G_{0}(z)-2\,i\pi\,G_{0}(z)\,G_{1}(u)+\frac{\pi^2}{3}\,G_{1}(u)
-\frac{\pi^2}{3}\,G_{1}(z)\nnb \\
-& 3\,i\pi\,G_{0}(z)\,G_{a_2}(u)+4\,i\pi\,G_{1}(z)\,G_{a_2}(u)
+\frac{3\pi^2}{2}\,G_{a_2}(u)-4\,G_{0}(z)\,G_{0,0}(u)\nnb \\
-& 2\,G_{1}(u)\,G_{0,0}(z)-2\,G_{a_2}(u)\,G_{0,0}(z)-3\,i\pi\,G_{0,0}(z)+2\,G_{1}(u)\,G_{0,1}(z)\nnb \\
+& 3\,G_{a_2}(u)\,G_{0,1}(z)+4\,i\pi\,G_{0,1}(z)+4\,G_{0}(z)\,G_{0,a_2}(u)-4\,G_{1}(z)\,G_{0,a_2}(u)
+4\,i\pi\,G_{0,a_2}(u)\nnb \\
+& 2\,G_{0}(z)\,G_{1,0}(u)+3\,G_{a_2}(u)\,G_{1,0}(z)+2\,i\pi\,G_{1,0}(z)-4\,G_{a_2}(u)\,G_{1,1}(z)+ 2\,G_{1,0,0}(z)\nnb \\
-& 2\,G_{0}(z)\,G_{1,a_2}(u)+2\,G_{1}(z)\,G_{1,a_2}(u)-2\,i\pi\,G_{1,a_2}(u)+3\,G_{0}(z)\,G_{a_2,0}(u)\nnb \\
-& 4\,G_{1}(z)\,G_{a_2,0}(u)+4\,i\pi\,G_{a_2,0}(u)-3\,G_{0}(z)\,G_{a_2,a_2}(u)+3\,G_{1}(z)\,G_{a_2,a_2}(u)\nnb \\
-& 3\,i\pi\,G_{a_2,a_2}(u)-2\,G_{0,0,0}(z)+3\,G_{0,0,1}(z)+3\,G_{0,1,0}(z)-4\,G_{0,1,1}(z)-4\,G_{0,a_2,0}(u)\nnb \\
-& 2\,G_{1,0,1}(z)+2\,G_{1,a_2,0}(u)-4\,G_{a_2,0,0}(u)+3\,G_{a_2,a_2,0}(u)-\zeta_3+\frac{1}{3}\,i\pi^3]+ {\cal O}(\ep^4) \, , \\[1.5em]
\tilde C_{5}=& \epcolor{\ep^3}\,[G_{a_1,0}(u)\,G_{0}(z)-\frac{\pi^2}{6}\,G_{0}(z)+\frac{\pi^2}{3}\,G_{a_1}(u)
+G_{a_1}(u)\,G_{0,1}(z)+G_{a_1}(u)\,G_{1,0}(z)\nnb \\
-& 2\,G_{1}(z)\,G_{a_1,a_1}(u)-G_{0,1,0}(z)-2\,G_{a_1,a_1,0}(u)+2\,\zeta_3]+ {\cal O}(\ep^4) \, , \\[1.5em]
\tilde C_{6}=& \epcolor{\ep^2}\,[-G_{0}(u)\,G_{0}(z)+G_{1}(z)\,G_{a_1}(u)-G_{0,1}(z)+G_{a_1,0}(u)-\frac{\pi^2}{6}]\nnb \\
+&\epcolor{\ep^3}\,[G_{0,0}(z)\,G_{0}(u)+4\,G_{0,1}(z)\,G_{0}(u)+2\,G_{1,0}(z)\,G_{0}(u)+\frac{\pi^2}{3}\,G_{0}(u)
+\frac{\pi^2}{2}\,G_{0}(z)\nnb \\
-&\frac{\pi^2}{3}\,G_{1}(u)+\frac{\pi^2}{3}\,G_{1}(z)-\frac{\pi^2}{2}\,G_{a_1}(u)
+4\,G_{0}(z)\,G_{0,0}(u)-2\,G_{1}(u)\,G_{0,1}(z)\nnb \\
-& G_{a_1}(u)\,G_{0,1}(z)-4\,G_{1}(z)\,G_{0,a_1}(u)-2\,G_{0}(z)\,G_{1,0}(u)-G_{a_1}(u)\,G_{1,0}(z)\nnb \\
-& 4\,G_{a_1}(u)\,G_{1,1}(z)+2\,G_{1}(z)\,G_{1,a_1}(u)-G_{0}(z)\,G_{a_1,0}(u)-4\,G_{1}(z)\,G_{a_1,0}(u)\nnb \\
+& 3\,G_{1}(z)\,G_{a_1,a_1}(u)+G_{0,0,1}(z)+G_{0,1,0}(z)+4\,G_{0,1,1}(z)-4\,G_{0,a_1,0}(u)+2\,G_{1,0,1}(z)\nnb \\
+& 2\,G_{1,a_1,0}(u)-4\,G_{a_1,0,0}(u)+3\,G_{a_1,a_1,0}(u)-\zeta_3]+ {\cal O}(\ep^4) \, , \\[1.5em]
\tilde C_{7}=& \epcolor{\ep}\,G_{0}(z)+\epcolor{\ep^2}\,[-G_{0,0}(z)-2\,G_{1,0}(z)+\frac{\pi^2}{3}]\nnb \\
+&\epcolor{\ep^3}\,[\frac{\pi^2}{3}\,G_{0}(z)-\frac{2\pi^2}{3}\,G_{1}(z)+G_{0,0,0}(z)+2\,G_{1,0,0}(z)+4\,G_{1,1,0}(z)-2\zeta_3]
+ {\cal O}(\ep^4) \, , \\[1.5em]
\tilde C_{8}=& \epcolor{\ep}\,[-G_{0}(u)-G_{1}(z)]\nnb \\
+&\epcolor{\ep^2}\,[4\,G_{0}(u)\,G_{1}(z)-G_{a_1}(u)\,G_{1}(z)+4\,G_{0,0}(u)+4\,G_{1,1}(z)-G_{a_1,0}(u)+\frac{\pi^2}{6}]\nnb \\
+&\epcolor{\ep^3}\,[-16\,G_{1,1}(z)\,G_{0}(u)-\frac{5\pi^2}{3}\,G_{0}(u)-\frac{5\pi^2}{3}\,G_{1}(z)+\frac{\pi^2}{6}\,G_{a_1}(u)
-16\,G_{1}(z)\,G_{0,0}(u)\nnb \\
+& 6\,G_{1}(z)\,G_{0,a_1}(u)+4\,G_{a_1}(u)\,G_{1,1}(z)+4\,G_{1}(z)\,G_{a_1,0}(u)-G_{1}(z)\,G_{a_1,a_1}(u)-7\,\zeta_3\nnb \\
-& 16\,G_{0,0,0}(u)+6\,G_{0,a_1,0}(u)-16\,G_{1,1,1}(z)+4\,G_{a_1,0,0}(u)-G_{a_1,a_1,0}(u)]
+ {\cal O}(\ep^4) \, , \\[1.5em]
\tilde C_{9}=&-1+\epcolor{\ep}\,[4\,G_{0}(u)+4\,G_{1}(z)]\nnb \\
+&\epcolor{\ep^2}\,[-16\,G_{0}(u)\,G_{1}(z)+6\,G_{a_1}(u)\,G_{1}(z)-16\,G_{0,0}(u)-16\,G_{1,1}(z)+6\,G_{a_1,0}(u)-\frac{5\pi^2}{3}]\nnb \\
+&\epcolor{\ep^3}\,[64\,G_{1,1}(z)\,G_{0}(u)+\frac{20\pi^2}{3}\,G_{0}(u)+\frac{20\pi^2}{3}\,G_{1}(z)-\pi^2\,G_{a_1}(u)+64\,G_{1}(z)\,G_{0,0}(u)\nnb \\
-&24\,G_{1}(z)\,G_{0,a_1}(u)-24\,G_{a_1}(u)\,G_{1,1}(z)-24\,G_{1}(z)\,G_{a_1,0}(u)+6\,G_{1}(z)\,G_{a_1,a_1}(u)+20\,\zeta_3\nnb \\
+&64\,G_{0,0,0}(u)-24\,G_{0,a_1,0}(u)+64\,G_{1,1,1}(z)-24\,G_{a_1,0,0}(u)+6\,G_{a_1,a_1,0}(u)]
+ {\cal O}(\ep^4) \, , \\[1.5em]
\tilde C_{10}=&\epcolor{\ep}\,[-G_{0}(u)+G_{0}(z)-G_{1}(z)+i\pi]\nnb \\
+&\epcolor{\ep^2}\,[-2\,G_{0}(z)\,G_{0}(u)+4\,G_{1}(z)\,G_{0}(u)-4\,i\pi\,G_{0}(u)+2\,i\pi\,G_{0}(z)-4\,i\pi\,G_{1}(z)\nnb \\
+&G_{0}(z)\,G_{a_2}(u)-G_{1}(z)\,G_{a_2}(u)+i\pi\,G_{a_2}(u)+4\,G_{0,0}(u)-2\,G_{0,1}(z)-2\,G_{1,0}(z)\nnb \\
+&4\,G_{1,1}(z)-G_{a_2,0}(u)-\frac{11\pi^2}{6}]\nnb \\
+&\epcolor{\ep^3}\,[-8\,i\pi\,G_{0}(z)\,G_{0}(u)+16\,i\pi\,G_{1}(z)\,G_{0}(u)-4\,G_{0,0}(z)\,G_{0}(u)+8\,G_{0,1}(z)\,G_{0}(u)\nnb \\
+& 8\,G_{1,0}(z)\,G_{0}(u)-16\,G_{1,1}(z)\,G_{0}(u)+\frac{19\pi^2}{3}\,G_{0}(u)-\frac{8}{3}\pi^2\,G_{0}(z)+\frac{19\pi^2}{3}\,G_{1}(z)\nnb \\
+& 2\,i\pi\,G_{0}(z)\,G_{a_2}(u)-4\,i\pi\,G_{1}(z)\,G_{a_2}(u) -\frac{11\pi^2}{6}\,G_{a_2}(u)+8\,G_{0}(z)\,G_{0,0}(u)\nnb \\
-& 16\,G_{1}(z)\,G_{0,0}(u)+16\,i\pi\,G_{0,0}(u)+4\,i\pi\,G_{0,0}(z)-2\,G_{a_2}(u)\,G_{0,1}(z)-8\,i\pi\,G_{0,1}(z)\nnb \\
-& 6\,G_{0}(z)\,G_{0,a_2}(u)+6\,G_{1}(z)\,G_{0,a_2}(u)-6\,i\pi\,G_{0,a_2}(u)-2\,G_{a_2}(u)\,G_{1,0}(z)-8\,i\pi\,G_{1,0}(z)\nnb \\
+& 4\,G_{a_2}(u)\,G_{1,1}(z)+16\,i\pi\,G_{1,1}(z)-2\,G_{0}(z)\,G_{a_2,0}(u)+4\,G_{1}(z)\,G_{a_2,0}(u)+4\,G_{0,0,0}(z)\nnb \\
-& 4\,i\pi\,G_{a_2,0}(u)+G_{0}(z)\,G_{a_2,a_2}(u)-G_{1}(z)\,G_{a_2,a_2}(u)+i\pi\,G_{a_2,a_2}(u)-16\,G_{0,0,0}(u)\nnb \\
-& 4\,G_{0,0,1}(z)-4\,G_{0,1,0}(z)+8\,G_{0,1,1}(z)+6\,G_{0,a_2,0}(u)-4\,G_{1,0,0}(z)+8\,G_{1,0,1}(z)\nnb \\
+& 8\,G_{1,1,0}(z)-16\,G_{1,1,1}(z)+4\,G_{a_2,0,0}(u)-G_{a_2,a_2,0}(u)-7\,\zeta_3-i\pi^3]
+ {\cal O}(\ep^4) \, , \\[1.5em]
\tilde C_{11}=&-1+\epcolor{\ep}\,[4\,G_{0}(u)-2\,G_{0}(z)+4\,G_{1}(z)-4\,i\pi]\nnb\\
+&\epcolor{\ep^2}\,[8\,G_{0}(z)\,G_{0}(u)-16\,G_{1}(z)\,G_{0}(u)+16\,i\pi\,G_{0}(u)-8\,i\pi\,G_{0}(z)+16\,i\pi\,G_{1}(z)\nnb\\
-& 6\,G_{0}(z)\,G_{a_2}(u)+6\,G_{1}(z)\,G_{a_2}(u)-6\,i\pi\,G_{a_2}(u)-16\,G_{0,0}(u)-4\,G_{0,0}(z)+8\,G_{0,1}(z)\nnb\\
+& 8\,G_{1,0}(z)-16\,G_{1,1}(z)+6\,G_{a_2,0}(u)+\frac{19\pi^2}{3}]\nnb\\
+& \epcolor{\ep^3}\,[32\,i\pi\,G_{0}(z)\,G_{0}(u)-64\,i\pi\,G_{1}(z)\,G_{0}(u)+16\,G_{0,0}(z)\,G_{0}(u)-32\,G_{0,1}(z)\,G_{0}(u)\nnb\\
-& 32\,G_{1,0}(z)\,G_{0}(u)+64\,G_{1,1}(z)\,G_{0}(u)-\frac{76\pi^2}{3}\,G_{0}(u)+\frac{38\pi^2}{3}\,G_{0}(z)-\frac{76\pi^2}{3}\,G_{1}(z)\nnb\\
-& 12\,i\pi\,G_{0}(z)\,G_{a_2}(u)+24\,i\pi\,G_{1}(z)\,G_{a_2}(u)+11\pi^2\,G_{a_2}(u)-32\,G_{0}(z)\,G_{0,0}(u)\nnb\\
+& 64\,G_{1}(z)\,G_{0,0}(u)-64\,i\pi\,G_{0,0}(u)-16\,i\pi\,G_{0,0}(z)+12\,G_{a_2}(u)\,G_{0,1}(z)+32\,i\pi\,G_{0,1}(z)\nnb\\
+& 24\,G_{0}(z)\,G_{0,a_2}(u)-24\,G_{1}(z)\,G_{0,a_2}(u)+24\,i\pi\,G_{0,a_2}(u)+12\,G_{a_2}(u)\,G_{1,0}(z)\nnb\\
+& 32\,i\pi\,G_{1,0}(z)-24\,G_{a_2}(u)\,G_{1,1}(z)-64\,i\pi\,G_{1,1}(z)+12\,G_{0}(z)\,G_{a_2,0}(u)+64\,G_{0,0,0}(u)\nnb\\
-& 24\,G_{1}(z)\,G_{a_2,0}(u)+24\,i\pi\,G_{a_2,0}(u)-6\,G_{0}(z)\,G_{a_2,a_2}(u)+6\,G_{1}(z)\,G_{a_2,a_2}(u)\nnb\\
-& 6\,i\pi\,G_{a_2,a_2}(u)-8\,G_{0,0,0}(z)+16\,G_{0,0,1}(z)+16\,G_{0,1,0}(z)-32\,G_{0,1,1}(z)\nnb\\
-& 24\,G_{0,a_2,0}(u)+16\,G_{1,0,0}(z)-32\,G_{1,0,1}(z)-32\,G_{1,1,0}(z)+64\,G_{1,1,1}(z)\nnb\\
-& 24\,G_{a_2,0,0}(u)+6\,G_{a_2,a_2,0}(u)+20\,\zeta_3+4\,i\pi^3]
+ {\cal O}(\ep^4) \, , \\[1.5em]
\tilde C_{12}=& 1 - \epcolor{\ep} \, G_{0}(z) +\epcolor{\ep^2}\,[G_{0,0}(z)+\frac{\pi^2}{3}]
+\epcolor{\ep^3}\,[-\frac{\pi^2}{3}\,G_{0}(z)-G_{0,0,0}(z)]
+ {\cal O}(\ep^4) \, .
\end{align}
}
\subsection{$C_{13}$ -- $C_{15}$}
\label{sec:C13}

The new integrals in this topology are $C_{13}$ -- $C_{15}$. However, in order to close the system of differential equations nine integrals are needed which we order as follows,
\begin{align}
\vec{C} =& \left\{\tilde C_{13},\tilde C_{5},\tilde C_{6},\tilde C_{7},\tilde C_{14},\tilde C_{8},\tilde C_{9},\tilde C_{15},\tilde C_{12}\right\}\,.
\end{align}
The corresponding matrix is $\tilde A_{13-15}$. The solution to the $C_{13}$ -- $C_{15}$ reads
\allowdisplaybreaks{
\begin{align}
\tilde C_{13}=& \epcolor{\ep^3}\,[G_{1}(z)\,G_{1,a_1}(u)+G_{1,a_1,0}(u)-G_{0,1}(z)\,G_{1}(u)-\frac{\pi^2}{6}\,G_{1}(u)-G_{0}(z)\,G_{1,0}(u)]
+ {\cal O}(\ep^4) \, , \\[1.5em]
\tilde C_{14}=& \epcolor{\ep^3}\,[\frac{\pi^2}{6}\,G_{0}(z)+G_{0,1,0}(z)-2\,\zeta_3]
+ {\cal O}(\ep^4) \, , \\[1.5em]
\tilde C_{15}=& -\frac{1}{4}-\epcolor{\ep^2}\,\frac{\pi^2}{4} -\epcolor{\ep^3}\,2\,\zeta_3 
+ {\cal O}(\ep^4) \, .
\end{align}
}
\subsection{$C_{16}$ -- $C_{22}$}
\label{sec:C16}

This topology has seven integrals, none of which has appeared in previous subsections. They are ordered according to
\begin{align}
\vec{C} =& \left\{\tilde C_{16},\tilde C_{17},\tilde C_{18},\tilde C_{19},\tilde C_{20},\tilde C_{21},\tilde C_{22}\right\}\,.
\end{align}
The corresponding matrix is $\tilde A_{16-22}$. The solution reads
\allowdisplaybreaks{
\begin{align}
\tilde C_{16}=& \epcolor{\ep^3}\,[G_{a_1}(u)\,G_{0,1}(z)-G_{a_2}(u)\,G_{0,1}(z)+G_{a_1}(u)\,G_{1,1}(z)
+G_{a_2}(u)\,G_{1,1}(z)\nnb\\
+& G_{1}(z)\,G_{a_1,0}(u)-G_{1}(z)\,G_{a_1,a_1}(u)+G_{1}(z)\,G_{a_2,0}(u)-2\,G_{0,0,1}(z)+G_{a_1,1,0}(u)\nnb\\
-& G_{a_1,a_1,0}(u)+G_{a_2,1,0}(u)-2\,\zeta_3]
+ {\cal O}(\ep^4) \, , \\[1.5em]
\tilde C_{17}=& \epcolor{\ep^2}\,[-G_{1}(z)\,G_{a_1}(u)+G_{0,1}(z)-G_{a_1,0}(u)+\frac{\pi^2}{6}]\nnb\\
+&\epcolor{\ep^3}\,[-2\,G_{0,1}(z)\,G_{1}(u)-\frac{\pi^2}{3}\,G_{1}(u)-\frac{\pi^2}{3}\,G_{1}(z)+\frac{\pi^2}{6}\,G_{a_1}(u)
+G_{a_1}(u)\,G_{0,1}(z)\nnb\\
+& G_{a_2}(u)\,G_{0,1}(z)+3\,G_{a_1}(u)\,G_{1,1}(z)-G_{a_2}(u)\,G_{1,1}(z)+2\,G_{1}(z)\,G_{1,a_1}(u)-G_{a_2,1,0}(u)\nnb\\
+& 3\,G_{1}(z)\,G_{a_1,0}(u)-2\,G_{1}(z)\,G_{a_1,a_1}(u)-G_{1}(z)\,G_{a_2,0}(u)+3\,G_{0,0,1}(z)-4\,G_{0,1,1}(z)\nnb\\
-& 2\,G_{1,0,1}(z)+2\,G_{1,a_1,0}(u)+2\,G_{a_1,0,0}(u)+G_{a_1,1,0}(u)-2\,G_{a_1,a_1,0}(u)+3\,\zeta_3]
+ {\cal O}(\ep^4) \, , \\[1.5em]
\tilde C_{18}=& \epcolor{\ep^2}\,[G_{0}(u)\,G_{1}(z)-G_{0,1}(z)+G_{1,0}(u)+G_{1,1}(z)]\nnb\\
+& \epcolor{\ep^3}\,[-G_{0,1}(z)\,G_{1}(u)+\frac{\pi^2}{6}\,G_{1}(u)+\frac{\pi^2}{6}\,G_{1}(z)
-2\,G_{1}(z)\,G_{0,0}(u)+G_{0}(u)\,G_{0,1}(z)\nnb\\
+& G_{1}(z)\,G_{0,a_1}(u)-4\,G_{1}(z)\,G_{1,0}(u)-6\,G_{0}(u)\,G_{1,1}(z)+G_{1}(z)\,G_{1,a_1}(u)-2\,G_{0,1,0}(u)\nnb\\
+& 5\,G_{0,1,1}(z)+G_{0,a_1,0}(u)-2\,G_{1,0,0}(u)-2\,G_{1,1,0}(u)-6\,G_{1,1,1}(z)+G_{1,a_1,0}(u)]
+ {\cal O}(\ep^4) \, , \\[1.5em]
\tilde C_{19}=& -\epcolor{\ep}\,G_{1}(z)+\epcolor{\ep^2}\,[4\,G_{1,1}(z)-G_{0,1}(z)]\nnb\\
+&\epcolor{\ep^3}\,[-\frac{2\pi^2}{3}\,G_{1}(z)-G_{0,0,1}(z)+4\,G_{0,1,1}(z)+6\,G_{1,0,1}(z)-16\,G_{1,1,1}(z)]
+ {\cal O}(\ep^4) \, , \\[1.5em]
\tilde C_{20}=&-1+\epcolor{\ep}\,4\,G_{1}(z)+\epcolor{\ep^2}\,[6\,G_{0,1}(z)-16\,G_{1,1}(z)-\frac{2\pi^2}{3}]\nnb\\
+&\epcolor{\ep^3}\,[\frac{8\pi^2}{3}\,G_{1}(z)+6\,G_{0,0,1}(z)-24\,G_{0,1,1}(z)-24\,G_{1,0,1}(z)+64\,G_{1,1,1}(z)+2\,\zeta_3]
+ {\cal O}(\ep^4) \, , \\[1.5em]
\tilde C_{21}=& \epcolor{\ep}\,[\,G_{0}(u)+G_{1}(z)]\nnb\\
+& \epcolor{\ep^2}\,[-2\,G_{0}(u)\,G_{1}(z)+G_{a_1}(u)\,G_{1}(z)-2\,G_{0,0}(u)-2\,G_{1,1}(z)+G_{a_1,0}(u)-\frac{\pi^2}{6}]\nnb\\
+& \epcolor{\ep^3}\,[4\,G_{1,1}(z)\,G_{0}(u)+\frac{2\pi^2}{3}\,G_{0}(u)+\frac{2\pi^2}{3}\,G_{1}(z)-\frac{\pi^2}{6}\,G_{a_1}(u)+4\,G_{1}(z)\,G_{0,0}(u)\nnb\\
-& 2\,G_{1}(z)\,G_{0,a_1}(u)-2\,G_{a_1}(u)\,G_{1,1}(z)-2\,G_{1}(z)\,G_{a_1,0}(u)+G_{1}(z)\,G_{a_1,a_1}(u)\nnb\\
+& 4\,G_{0,0,0}(u)-2\,G_{0,a_1,0}(u)+4\,G_{1,1,1}(z)-2\,G_{a_1,0,0}(u)+G_{a_1,a_1,0}(u)+\zeta_3]
+ {\cal O}(\ep^4) \, , \\[1.5em]
\tilde C_{22}=& 1 + \epcolor{\ep^2}\,\frac{\pi^2}{3} + {\cal O}(\ep^4) \, .
\end{align}
}
\subsection{$C_{23}$ -- $C_{27}$}
\label{sec:C23}

This topology has also seven integrals, of which $C_{23}$ -- $C_{27}$ are new. The entire topology reads
\begin{align}
\vec{C} =& \left\{\tilde C_{23},\tilde C_{24},\tilde C_{25},\tilde C_{7},\tilde C_{26},\tilde C_{27},\tilde C_{12}\right\}\,.
\end{align}
The corresponding matrix is $\tilde A_{23-27}$. The solution reads
\allowdisplaybreaks{
\begin{align}
\tilde C_{23}=& \epcolor{\ep^3}\,[-G_{a_2,0}(u)\,G_{0}(z)+\frac{\pi^2}{2}\,G_{0}(z)+\frac{\pi^2}{6}\,G_{a_1}(u)+\frac{\pi^2}{2}\,G_{a_2}(u)
+G_{a_2}(u)\,G_{0,0}(z)\nnb\\
-& G_{a_2}(u)\,G_{0,1}(z)-G_{a_2}(u)\,G_{1,0}(z)+G_{a_1}(u)\,G_{1,1}(z)+G_{a_2}(u)\,G_{1,1}(z)+G_{1}(z)\,G_{a_1,0}(u)\nnb\\
-& G_{1}(z)\,G_{a_1,a_1}(u)+G_{1}(z)\,G_{a_2,0}(u)+G_{0,0,0}(z)-G_{0,1,0}(z)+G_{a_1,0,0}(u)-G_{a_1,a_1,0}(u)\nnb\\
+& G_{a_2,0,0}(u)+2\,\zeta_3]
+ {\cal O}(\ep^4) \, , \\[1.5em]
\tilde C_{24}=& \epcolor{\ep^2}\,[\,G_{0}(u)\,G_{0}(z)-G_{1}(z)\,G_{a_1}(u)+G_{0,1}(z)-G_{a_1,0}(u)+\frac{\pi^2}{6}]\nnb\\
+& \epcolor{\ep^3}\,[-G_{0,0}(z)\,G_{0}(u)-4\,G_{0,1}(z)\,G_{0}(u)-2\,G_{1,0}(z)\,G_{0}(u)-\frac{\pi^2}{3}\,G_{0}(u)-\pi^2\,G_{0}(z)\nnb\\
+& \frac{\pi^2}{3}\,G_{1}(u)-\frac{\pi^2}{3}\,G_{1}(z)+\frac{\pi^2}{3}\,G_{a_1}(u)-\frac{\pi^2}{2}\,G_{a_2}(u)
-4\,G_{0}(z)\,G_{0,0}(u)-G_{a_2}(u)\,G_{0,0}(z)\nnb\\
+& 2\,G_{1}(u)\,G_{0,1}(z)+G_{a_1}(u)\,G_{0,1}(z)+G_{a_2}(u)\,G_{0,1}(z)+4\,G_{1}(z)\,G_{0,a_1}(u)-2\,G_{a_1,a_1,0}(u)\nnb\\
+& 2\,G_{0}(z)\,G_{1,0}(u)+G_{a_1}(u)\,G_{1,0}(z)+G_{a_2}(u)\,G_{1,0}(z)+3\,G_{a_1}(u)\,G_{1,1}(z)-G_{a_2,0,0}(u)\nnb\\
-& G_{a_2}(u)\,G_{1,1}(z)-2\,G_{1}(z)\,G_{1,a_1}(u)+G_{0}(z)\,G_{a_1,0}(u)+3\,G_{1}(z)\,G_{a_1,0}(u)+3\,G_{a_1,0,0}(u)\nnb\\
-& 2\,G_{1}(z)\,G_{a_1,a_1}(u)+G_{0}(z)\,G_{a_2,0}(u)-G_{1}(z)\,G_{a_2,0}(u)-G_{0,0,0}(z)-G_{0,0,1}(z)\nnb\\
-& 4\,G_{0,1,1}(z)+4\,G_{0,a_1,0}(u)-2\,G_{1,0,1}(z)-2\,G_{1,a_1,0}(u)-\zeta_3]
+ {\cal O}(\ep^4) \, , \\[1.5em]
\tilde C_{25}=&\epcolor{\ep^2}\,[G_{0}(u)\,(G_{1}(z)-G_{0}(z))+G_{0,0}(u)+G_{0,0}(z)-G_{0,1}(z)-G_{1,0}(z)+G_{1,1}(z)+\frac{\pi^2}{2}]\nnb\\
+& \epcolor{\ep^3}\,[3\,G_{0,1}(z)\,G_{0}(u)+2\,G_{1,0}(z)\,G_{0}(u)-6\,G_{1,1}(z)\,G_{0}(u)-\pi^2\,G_{0}(u)-\frac{\pi^2}{2}\,G_{0}(z)\nnb\\
-& \frac{\pi^2}{6}\,G_{1}(u)-\frac{7\pi^2}{6}\,G_{1}(z)+3\,G_{0}(z)\,G_{0,0}(u)-6\,G_{1}(z)\,G_{0,0}(u)-G_{1}(u)\,G_{0,1}(z)\nnb\\
+& G_{1}(z)\,G_{0,a_1}(u)-G_{0}(z)\,G_{1,0}(u)+G_{1}(z)\,G_{1,a_1}(u)-6\,G_{0,0,0}(u)-3\,G_{0,0,0}(z)\nnb\\
+& G_{0,1,0}(z)+3\,G_{0,1,1}(z)+G_{0,a_1,0}(u)+2\,G_{1,0,1}(z)+2\,G_{1,1,0}(z)-6\,G_{1,1,1}(z)\nnb\\
+& G_{1,a_1,0}(u)]+ {\cal O}(\ep^4) \, , \\[1.5em]
\tilde C_{26}=&\epcolor{\ep}\,[G_{0}(u)+G_{1}(z)]\nnb\\
+& \epcolor{\ep^2}\,[-G_{0}(u)\,G_{0}(z)-2\,G_{0}(u)\,G_{1}(z)+G_{1}(z)\,G_{a_1}(u)-2\,G_{0,0}(u)-G_{0,1}(z)-G_{1,0}(z)\nnb\\
-& 2\,G_{1,1}(z)+G_{a_1,0}(u)-\frac{\pi^2}{6}]\nnb\\
+& \epcolor{\ep^3}\,[G_{0,0}(z)\,G_{0}(u)+2\,G_{0,1}(z)\,G_{0}(u)+2\,G_{1,0}(z)\,G_{0}(u)+4\,G_{1,1}(z)\,G_{0}(u)
+\frac{2\pi^2}{3}\,G_{0}(u)\nnb\\
+& \frac{\pi^2}{6}\,G_{0}(z)+\frac{2\pi^2}{3}\,G_{1}(z)-\frac{\pi^2}{6}\,G_{a_1}(u)+2\,G_{0}(z)\,G_{0,0}(u)+4\,G_{1}(z)\,G_{0,0}(u)+G_{1,0,0}(z)\nnb\\
-& G_{a_1}(u)\,G_{0,1}(z)-2\,G_{1}(z)\,G_{0,a_1}(u)-G_{a_1}(u)\,G_{1,0}(z)-2\,G_{a_1}(u)\,G_{1,1}(z)+2\,G_{1,0,1}(z)\nnb\\
-& G_{0}(z)\,G_{a_1,0}(u)-2\,G_{1}(z)\,G_{a_1,0}(u)+G_{1}(z)\,G_{a_1,a_1}(u)+4\,G_{0,0,0}(u)+G_{0,0,1}(z)\nnb\\
+& G_{0,1,0}(z)+2\,G_{0,1,1}(z)-2\,G_{0,a_1,0}(u)+2\,G_{1,1,0}(z)+4\,G_{1,1,1}(z)-2\,G_{a_1,0,0}(u)\nnb\\
+& G_{a_1,a_1,0}(u)+\zeta_3]
+ {\cal O}(\ep^4) \, , \\[1.5em]
\tilde C_{27}=&-\frac{1}{4}+\epcolor{\ep}\,\frac{1}{2}\,G_{0}(z)+\epcolor{\ep^2}\,[-G_{0,0}(z)-\frac{\pi^2}{4}]
+ \epcolor{\ep^3}\,[\frac{\pi^2}{2}\,G_{0}(z)+2\,G_{0,0,0}(z)-2\,\zeta_3]
+ {\cal O}(\ep^4) \, .
\end{align}
}
\subsection{$C_{28}$ -- $C_{32}$}
\label{sec:C28}

This topology has ten integrals, of which the five integrals $C_{28}$ -- $C_{32}$ are new. They are embedded in the topology as follows
\begin{align}
\vec{C} =& \left\{\tilde C_{28},\tilde C_{29},\tilde C_{30},\tilde C_{7},\tilde C_{26},\tilde C_{21},\tilde C_{22},\tilde C_{31},\tilde C_{32},\tilde C_{12}\right\}\,.
\end{align}
The corresponding matrix is $\tilde A_{28-32}$. The solution reads
\allowdisplaybreaks{
\begin{align}
\tilde C_{28}=& \epcolor{\ep^3}\,[-G_{1,0}(u)\,G_{0}(z)+\frac{\pi^2}{6}\,G_{0}(z)-\frac{\pi^2}{6}\,G_{1}(u)-\frac{\pi^2}{6}\,G_{a_1}(u)
-G_{1}(u)\,G_{0,1}(z)\nnb\\
-& G_{1}(z)\,G_{0,a_1}(u)+G_{1}(z)\,G_{1,a_1}(u)+G_{1}(z)\,G_{a_1,a_1}(u)-G_{0,a_1,0}(u)+G_{1,a_1,0}(u)\nnb\\
+& G_{a_1,a_1,0}(u)]+ {\cal O}(\ep^4) \, , \\[1.5em]
\tilde C_{29}=& \epcolor{\ep^3}\,[G_{0,1}(z)\,G_{1}(u)+\frac{\pi^2}{6}\,G_{1}(u)-\frac{\pi^2}{6}\,G_{a_1}(u)-G_{a_1}(u)\,G_{0,1}(z)
+G_{1}(z)\,G_{0,a_1}(u)\nnb\\
+& G_{0}(z)\,G_{1,0}(u)-G_{a_1}(u)\,G_{1,0}(z)-G_{1}(z)\,G_{1,a_1}(u)-G_{0}(z)\,G_{a_1,0}(u)+G_{1}(z)\,G_{a_1,a_1}(u)\nnb\\
+& G_{0,1,0}(z)+G_{0,a_1,0}(u)-G_{1,a_1,0}(u)+G_{a_1,a_1,0}(u)-2\,\zeta_3]
+ {\cal O}(\ep^4) \, , \\[1.5em]
\tilde C_{30}=& \epcolor{\ep^2}\,[-G_{0}(u)\,G_{0}(z)+G_{1}(z)\,G_{a_1}(u)-G_{0,1}(z)+G_{a_1,0}(u)-\frac{\pi^2}{6}]\nnb\\
+& \epcolor{\ep^3}\,[2\,G_{0,0}(u)\,G_{0}(z)-3\,G_{1,0}(u)\,G_{0}(z)-2\,G_{a_1,0}(u)\,G_{0}(z)+2\,G_{a_3,0}(u)\,G_{0}(z)\nnb\\
+& 2\,G_{a_4,0}(u)\,G_{0}(z)+\frac{\pi^2}{6}\,G_{0}(z)-\frac{\pi^2}{2}\,G_{1}(u)+\frac{\pi^2}{3}\,G_{1}(z)
-\frac{2\pi^2}{3}\,G_{a_1}(u)+\frac{2\pi^2}{3}\,G_{a_3}(u)\nnb\\
+& \frac{2\pi^2}{3}\,G_{a_4}(u)-2\,G_{a_3}(u)\,G_{-1,0}(\sqrt{z})+2\,G_{a_4}(u)\,G_{-1,0}(\sqrt{z})+G_{0}(u)\,G_{0,0}(z)-G_{0,a_1,0}(u)\nnb\\
+& 2\,G_{0}(u)\,G_{0,1}(z)-3\,G_{1}(u)\,G_{0,1}(z)-2\,G_{a_1}(u)\,G_{0,1}(z)+2\,G_{a_3}(u)\,G_{0,1}(z)+G_{0,1,0}(z)\nnb\\
+& 2\,G_{a_4}(u)\,G_{0,1}(z)-G_{1}(z)\,G_{0,a_1}(u)+2\,G_{a_3}(u)\,G_{1,0}(\sqrt{z})-2\,G_{a_4}(u)\,G_{1,0}(\sqrt{z})\nnb\\
+& 2\,G_{0}(u)\,G_{1,0}(z)-2\,G_{a_1}(u)\,G_{1,0}(z)+G_{a_3}(u)\,G_{1,0}(z)+G_{a_4}(u)\,G_{1,0}(z)+G_{0,0,1}(z)\nnb\\
-& 2\,G_{a_1}(u)\,G_{1,1}(z)+3\,G_{1}(z)\,G_{1,a_1}(u)-2\,G_{1}(z)\,G_{a_1,0}(u)+4\,G_{1}(z)\,G_{a_1,a_1}(u)+\zeta_3\nnb\\
-& 4\,G_{1}(z)\,G_{a_3,a_1}(u)-4\,G_{1}(z)\,G_{a_4,a_1}(u)+2\,G_{0,1,1}(z)+2\,G_{1,0,1}(z)-4\,G_{a_4,a_1,0}(u)\nnb\\
+& 3\,G_{1,a_1,0}(u)-2\,G_{a_1,0,0}(u)+4\,G_{a_1,a_1,0}(u)
-4\,G_{a_3,a_1,0}(u)]
+ {\cal O}(\ep^4) \, , \\[1.5em]
\tilde C_{31}=&-\epcolor{\ep^2}\,\frac{1}{2}\,G_{1,0}(z)
+\epcolor{\ep^3}\,[-G_{-1,-1,0}(\sqrt{z})+G_{-1,1,0}(\sqrt{z})-\frac{1}{2}\,G_{0,1,0}(z)+G_{1,-1,0}(\sqrt{z})\nnb\\
+& \frac{1}{2}\,G_{1,0,0}(z)-G_{1,1,0}(\sqrt{z})+\frac{3}{2}\,G_{1,1,0}(z)]
+ {\cal O}(\ep^4) \, , \\[1.5em]
\tilde C_{32}=& \epcolor{\ep^2}\,[G_{-1,0}(\sqrt{z})-G_{1,0}(\sqrt{z})]\nnb\\
+& \epcolor{\ep^3}\,[-2\,G_{-1,-1,0}(\sqrt{z})-2\,G_{-1,0,0}(\sqrt{z})-4\,G_{-1,1,0}(\sqrt{z})
-2\,G_{0,-1,0}(\sqrt{z})\nnb\\
+& 2\,G_{0,1,0}(\sqrt{z})+4\,G_{1,-1,0}(\sqrt{z})+2\,G_{1,0,0}(\sqrt{z})+2\,G_{1,1,0}(\sqrt{z})]
+ {\cal O}(\ep^4) \, .
\end{align}
}
\subsection{$C_{33}$ and $C_{34}$}
\label{sec:C33}

This topology has seven integrals, of which only $C_{33}$ -- $C_{34}$ have not yet appeared in the previous topologies. The integrals are ordered as
\begin{align}
\vec{C} =& \left\{\tilde C_{33},\tilde C_{34},\tilde C_{7},\tilde C_{22},\tilde C_{31},\tilde C_{32},\tilde C_{12}\right\}\,.
\end{align}
The corresponding matrix is $\tilde A_{33,34}$. The solution to ${\cal O}(\ep^3)$ is very short
\allowdisplaybreaks{
\begin{align}
\tilde C_{33}=& \epcolor{\ep^3}\,[G_{0,1,0}(z)-2\,\zeta_3] + {\cal O}(\ep^4) \, , \\[1.5em]
\tilde C_{34}=& \epcolor{\ep^3}\,\frac{\pi^2}{6}\,G_{0}(z) + {\cal O}(\ep^4) \, .
\end{align}
}
At order ${\cal O}(\ep^4)$ the solution requires also Goncharov polylogarithms of argument $\sqrt{z}$.
\subsection{$C_{35}$}
\label{sec:C35}

This topology has three integrals, and only $C_{35}$ is new. The integrals are ordered as
\begin{align}
\vec{C} =& \left\{\tilde C_{35},\tilde C_{19},\tilde C_{20}\right\}\,.
\end{align}
The corresponding matrix is $\tilde A_{35}$. The solution reads
\allowdisplaybreaks{
\begin{align}
\tilde C_{35}=& \epcolor{\ep}\,\frac{1}{2}\,G_{0}(u)+\epcolor{\ep^2}\,[-2\,G_{0}(u)\,G_{1}(z)+G_{a_1}(u)\,G_{1}(z)-\frac{3}{2}\,G_{0,0}(u)-G_{0,1}(z)-\frac{1}{2}\,G_{1,0}(u)
\nnb\\
+& G_{a_1,0}(u)-\frac{\pi^2}{12}]\nnb\\
+& \epcolor{\ep^3}\,[8\,G_{1,1}(z)\,G_{0}(u)+\frac{7\pi^2}{12}\,G_{0}(u)+\frac{\pi^2}{12}\,G_{1}(u)+\frac{\pi^2}{3}\,G_{1}(z)
-\frac{\pi^2}{6}\,G_{a_1}(u)+\frac{9}{2}\,G_{0,0,0}(u)\nnb\\
+& 6\,G_{1}(z)\,G_{0,0}(u)+G_{1}(u)\,G_{0,1}(z)-G_{a_1}(u)\,G_{0,1}(z)-3\,G_{1}(z)\,G_{0,a_1}(u)-3\,G_{0,a_1,0}(u)\nnb\\
+& 2\,G_{1}(z)\,G_{1,0}(u)-4\,G_{a_1}(u)\,G_{1,1}(z)-G_{1}(z)\,G_{1,a_1}(u)-4\,G_{1}(z)\,G_{a_1,0}(u)+4\,G_{1,0,1}(z)\nnb\\
+& 2\,G_{1}(z)\,G_{a_1,a_1}(u)-G_{0,0,1}(z)+\frac{3}{2}\,G_{0,1,0}(u)+4\,G_{0,1,1}(z)+\frac{3}{2}\,G_{1,0,0}(u)
+\frac{1}{2}\,G_{1,1,0}(u)\nnb\\
-& G_{1,a_1,0}(u)-3\,G_{a_1,0,0}(u)-G_{a_1,1,0}(u)+2\,G_{a_1,a_1,0}(u)+\zeta_3]
+ {\cal O}(\ep^4) \, .
\end{align}
}
\subsection{$C_{36}$ and $C_{37}$}
\label{sec:C36}

This topology has four integrals, of which $C_{36}$ and $C_{37}$ are new. The integrals are
\begin{align}
\vec{C} =& \left\{\tilde C_{36},\tilde C_{37},\tilde C_{26},\tilde C_{12}\right\}\,.
\end{align}
The corresponding matrix is $\tilde A_{36,37}$. The solution reads
\allowdisplaybreaks{
\begin{align}
\tilde C_{36}=& \epcolor{\ep^2}\,[-G_{0}(z)\,G_{0}(u)+G_{1}(z)\,G_{0}(u)-i\pi\,G_{0}(u)+G_{1}(u)\,G_{1}(z)-i\pi\,G_{1}(z)\nnb\\
+& G_{0,1}(u)-G_{0,1}(z)+G_{1,0}(u)-G_{1,0}(z)+2\,G_{1,1}(z)]\nnb\\
+& \epcolor{\ep^3}\,[4\,i\pi\,G_{1}(z)\,G_{0}(u)+G_{0,0}(z)\,G_{0}(u)+2\,G_{0,1}(z)\,G_{0}(u)+3\,G_{1,0}(z)\,G_{0}(u)\nnb\\
-& 6\,G_{1,1}(z)\,G_{0}(u)+\frac{2\pi^2}{3}\,G_{0}(u)+\frac{\pi^2}{6}\,G_{0}(z)-\frac{\pi^2}{6}\,G_{1}(u)
+2\,i\pi\,G_{1}(u)\,G_{1}(z)+\frac{\pi^2}{2}\,G_{1}(z)\nnb\\
-& 2\,i\pi\,G_{1}(z)\,G_{a_1}(u)+2\,G_{0}(z)\,G_{0,0}(u)-2\,G_{1}(z)\,G_{0,0}(u)+2\,i\pi\,G_{0,0}(u)+G_{0}(z)\,G_{0,1}(u)\nnb\\
-& 4\,G_{1}(z)\,G_{0,1}(u)+2\,i\pi\,G_{0,1}(u)+G_{1}(u)\,G_{0,1}(z)-2\,G_{a_1}(u)\,G_{0,1}(z)-G_{0}(z)\,G_{0,a_1}(u)\nnb\\
+& G_{1}(z)\,G_{0,a_1}(u)-i\pi\,G_{0,a_1}(u)+G_{0}(z)\,G_{1,0}(u)-4\,G_{1}(z)\,G_{1,0}(u)+2\,i\pi\,G_{1,0}(u)\nnb\\
+& G_{1}(u)\,G_{1,0}(z)-2\,G_{a_1}(u)\,G_{1,0}(z)-2\,G_{1}(z)\,G_{1,1}(u)-6\,G_{1}(u)\,G_{1,1}(z)+6\,i\pi\,G_{1,1}(z)\nnb\\
+& 4\,G_{a_1}(u)\,G_{1,1}(z)+G_{1}(z)\,G_{1,a_1}(u)-2\,G_{0}(z)\,G_{a_1,0}(u)+2\,G_{1}(z)\,G_{a_1,0}(u)+3\,G_{1,0,1}(z)\nnb\\
-& 2\,i\pi\,G_{a_1,0}(u)+2\,G_{1}(z)\,G_{a_1,1}(u)-2\,G_{0,0,1}(u)+G_{0,0,1}(z)-2\,G_{0,1,0}(u)-2\,G_{1,0,1}(u)\nnb\\
-& 2\,G_{0,1,1}(u)+2\,G_{0,1,1}(z)+G_{0,a_1,1}(u)-2\,G_{1,0,0}(u)+G_{1,0,0}(z)+G_{0,1,0}(z)-2\,G_{1,1,0}(u)\nnb\\
+& 4\,G_{1,1,0}(z)-12\,G_{1,1,1}(z)+G_{1,a_1,0}(u)+2\,G_{a_1,0,1}(u)+2\,G_{a_1,1,0}(u)+\frac{1}{6}\,i\pi^3]
+ {\cal O}(\ep^4) \, , \\[1.5em]
\tilde C_{37}=& \epcolor{\ep}\,[-G_{0}(z)+G_{1}(u)+G_{1}(z)-i\pi]\nnb\\
+& \epcolor{\ep^2}\,[G_{0}(z)\,G_{1}(u)-2\,G_{1}(z)\,G_{1}(u)+2\,i\pi\,G_{1}(u)+2\,i\pi\,G_{1}(z)
-G_{0}(z)\,G_{a_1}(u)+G_{a_1,1}(u)\nnb\\
+& G_{1}(z)\,G_{a_1}(u)-i\pi\,G_{a_1}(u)+G_{0,0}(z)+G_{1,0}(z)-2\,G_{1,1}(u)-2\,G_{1,1}(z)+\frac{2\pi^2}{3}]\nnb\\
+& \epcolor{\ep^3}\,[-2\,G_{1,1}(u)\,G_{0}(z)+2\,G_{1,a_1}(u)\,G_{0}(z)+G_{a_1,1}(u)\,G_{0}(z)-G_{a_1,a_1}(u)\,G_{0}(z)\nnb\\
-& \frac{\pi^2}{3}\,G_{0}(z)-\pi^2\,G_{1}(u)-4\,i\pi\,G_{1}(u)\,G_{1}(z)-\pi^2\,G_{1}(z)+2\,i\pi\,G_{1}(z)\,G_{a_1}(u)+4\,G_{1,1,1}(z)\nnb\\
+& \frac{2\pi^2}{3}\,G_{a_1}(u)-G_{1}(u)\,G_{0,0}(z)+G_{a_1}(u)\,G_{0,0}(z)-2\,G_{1}(u)\,G_{1,0}(z)+G_{a_1}(u)\,G_{1,0}(z)\nnb\\
+& 4\,G_{1}(z)\,G_{1,1}(u)-4\,i\pi\,G_{1,1}(u)+4\,G_{1}(u)\,G_{1,1}(z)-2\,G_{a_1}(u)\,G_{1,1}(z)-4\,i\pi\,G_{1,1}(z)\nnb\\
-& 2\,G_{1}(z)\,G_{1,a_1}(u)+2\,i\pi\,G_{1,a_1}(u)-2\,G_{1}(z)\,G_{a_1,1}(u)+2\,i\pi\,G_{a_1,1}(u)+G_{1}(z)\,G_{a_1,a_1}(u)\nnb\\
-& i\pi\,G_{a_1,a_1}(u)-G_{0,0,0}(z)-G_{1,0,0}(z)-2\,G_{1,1,0}(z)+4\,G_{1,1,1}(u)-2\,G_{1,a_1,1}(u)\nnb\\
-& 2\,G_{a_1,1,1}(u)+G_{a_1,a_1,1}(u)+2\,\zeta_3]
+ {\cal O}(\ep^4) \, .
\end{align}
}
\subsection{$C_{38}$ and $C_{39}$}
\label{sec:C38}

These integrals arise from diagrams with a massive quark loop inside a gluon propagator. They appeared in a slightly different version already in the calculation of the two-loop tree amplitudes in $B\to \pi\pi$~\cite{Bell:2009nk,Beneke:2009ek}, and analytic results were recently derived in~\cite{Bell:2014zya} as $M_{28,29}$. It turns out that the results of $C_{38,39}$ can be obtained from the latter reference if one adjusts the kinematics to the present problem. To be precise, one has to replace
\begin{align}
u \to u \, (1-z)
\end{align}
in the expressions in~\cite{Bell:2014zya}. That is, in the definition of the canonical basis (cf.~(3.30) and (3.31) of~\cite{Bell:2014zya} and~(\ref{eq:C38}),~(\ref{eq:C39}) of the present article), and also in the solution, eqs.~(4.64) and~(4.65) of~\cite{Bell:2014zya}. In particular, the kinematic variable $p$ changes to ($\bar z=1-z$)
\begin{align}
p=& \frac{1-\sqrt{(2-u\bar z)^2-4\bar z(1-u\bar z)}}{1-u\bar z} \; .
\end{align}


\section{Checks}
\label{sec:checks}

In order to validate the analytic results presented above, we performed several checks of analytic and numeric nature. Those integrals that possess a closed form in terms of hypergeometric functions were analytically expanded in $\eps$ using {\tt HypExp}~\cite{Huber:2005yg,Huber:2007dx}. Subsequently, we re-wrote the resulting polylogarithms and HPLs in terms of Goncharov polylogarithms and compared to the results obtained by the differential equation method.

For the numerical checks we used a dozen points in the $u-z$ plane. We first evaluated the Goncharov polylogarithms that appear in our analytic results numerically with the GiNaC-library~\cite{Bauer:2000cp,Vollinga:2004sn}. We also derived Mellin-Barnes (MB) representations, partially using the {\tt AMBRE}-package~\cite{Gluza:2007rt}. The analytic continuation to $\eps=0$ and subsequent numerical integration was carried out by {\tt MB.m}~\cite{Czakon:2005rk}. This worked for almost all cases, even in the presence of kinematic thresholds, and yielded agreement to the GiNaC results to $5 \cdot 10^{-10}$ or better. There are, however, a few cases in which the Monte-Carlo integration implemented in {\tt MB.m} failed due to highly oscillating integrands, notably for the integrals $C_{28-30}$, and their ``mass-flipped'' counterparts (where $m_c \leftrightarrow m_b$ and $q_3 \leftrightarrow q_4$). In these cases, we relied on the sector decomposition method implemented in {\tt SecDec}~\cite{Carter:2010hi,Borowka:2012yc}, which yielded agreement with GiNaC at the level of $8 \cdot 10^{-7}$ for the highest $\eps$-coefficients in $C_{28-30}$, and at the level of $6 \cdot 10^{-4}$ for the highest $\eps$-coefficients of their ``mass-flipped'' counterparts. The agreement is several orders of magnitude better for the lower coefficients in the $\eps$-expansion.

Another important point to mention is the fact that the GiNaC results were obtained in the canonical basis, whereas most of the MB representations and the {\tt SecDec} results were derived in an ``ordinary'' basis of un-dotted and singly-dotted master integrals. The change of basis was then performed using the Laporta reduction. Having calculated the numerics in two different integral bases constitutes another non-trivial check of our results.


\section{Conclusion}
\label{sec:conclusion}

We obtained analytic results to all two-loop master integrals that are necessary for the description of the non-leptonic decay $B\to D\pi$ at NNLO in QCD factorisation. They are expressed in terms of Goncharov polylogarithms of argument $u$ and weights that are either integer numbers ($0$ or~$\pm 1$), or contain the second kinematic varible, $z$. It is remarkable that six $z$-dependent weights are sufficient for writing down the entire set of solutions, including the ``mass-flipped'' integrals.

With the master integrals at hand, the bare two-loop part of the hard-scattering kernels $T_{ij}(u)$ in~(\ref{eq:bbns}) is complete. The remaining task consists of renormalising the ultraviolet divergences and subtracting infrared divergences via matching from QCD onto soft-collinear effective theory. Steps towards this goal are outlined in~\cite{Huber:2014kaa}. Having the hard-scattering kernels $T_{ij}(u)$ written in terms of iterated integrals is an optimal choice for carrying out the convolution integral with the pion LCDA in~(\ref{eq:bbns}), and it might be feasible to obtain the NNLO topological tree amplitude in analytic form. In any case our results constitute an important step towards the phenomenology of $B\rightarrow D\pi$ decays at NNLO in QCD factorisation.

Let us compare the integrals in the present work to those recently obtained in~\cite{Bell:2014zya} during the evaluation of the two-loop penguin amplitude. Both are two-loop problems with scales $u$ and $z$. The present integrals are a bit less involved compared to those in ~\cite{Bell:2014zya}, in a sense that the linear combinations that form a canonical master integral are shorter, the occurring weights are fewer, and the choice of kinematic invariants is less complicated. The main reason for this is that in the present work the external kinematics of the final state contains also the second internal mass, notably $m_c$. On the other hand, the only five-line integral in~\cite{Bell:2014zya}, a two-point function ($M_{22}$), is in fact a one-scale integral, whereas here we encountered several five-line integrals with four external legs which are genuine two-scale functions. Moreover, most of our integrals are needed to order ${\cal O}(\ep^4)$, whereas in~\cite{Bell:2014zya} all but two integrals were required only to order ${\cal O}(\ep^3)$.

On more general grounds, it will be interesting to investigate how the canonical basis depends on the number of loops, legs, scales, space-time dimensions, and on the external kinematics. Every example therefore sharpens our understanding of the patterns that such bases follow, with the goal of eventually developing an algorithm for their automated construction.


\section*{Acknowledgments}
We would like to thank A.~Smirnov for useful correspondence on {\tt FIRE}~\cite{Smirnov:2008iw}. This work is supported by DFG research unit FOR 1873 ``Quark Flavour Physics and Effective Field Theories''.

\appendix


\section{The matrices $\tilde A$}
\label{app:Atilde}

Here we list the matrices $\tilde A$ for the different topologies. Their entries can all be expressed in terms of the following nine logarithms,
\allowdisplaybreaks{
\begin{align}
L_1 =& \ln(u) \; , & L_6 =& \ln(z+u(1-z)) \; , \nnb \\
L_2 =& \ln(1-u) \; , & L_7 =& \ln\left(1-u \left(1-\sqrt{z}\right)\right) \; , \nnb \\
L_3 =& \ln(z) \; , & L_8 =& \ln\left(1-u \left(1+\sqrt{z}\right)\right) \; , \nnb  \\
L_4 =& \ln(1-z) \; , & L_9 =& \ln\left(\frac{1-\sqrt{z}}{1+\sqrt{z}}\right) \; .  \\
L_5 =& \ln(1-u(1-z))\; , & \nnb 
\end{align}}
The matrices $\tilde A$ now assume the following compact form,

\allowdisplaybreaks{
\begin{align}
\hspace*{-18pt} \tilde A_{1-12} = &
\scalebox{0.8}{\parbox{16cm}{$\dps\left(
\begin{array}{ccccc}
 -4 L_1-L_4 & 3 L_3-3 L_4 & -2 L_2-\frac{L_3}{2}-\frac{L_4}{2}+L_6 & -L_2+\frac{L_3}{2}-L_4+\frac{L_6}{2} & L_2+\frac{L_3}{2}-\frac{L_4}{2} \\
 -3 L_4 & -4 L_1-L_3-L_4 & L_2-\frac{L_4}{2} & -L_2-L_4 & -2 L_2-\frac{L_4}{2}+L_5 \\
 0 & 0 & 2 L_2-L_3+2 L_4-2 L_6 & L_3-L_6 & 0 \\
 0 & 0 & -2 L_1-2 L_4+2 L_6 & -4 L_1+2 L_2-2 L_4+L_6 & 0 \\
 0 & 0 & 0 & 0 & 2 L_2-L_3+2 L_4-2 L_5 \\
 0 & 0 & 0 & 0 & -2 L_1-2 L_4+2 L_5 \\
 0 & 0 & 0 & 0 & 0  \\
 0 & 0 & 0 & 0 & 0  \\
 0 & 0 & 0 & 0 & 0  \\
 0 & 0 & 0 & 0 & 0  \\
 0 & 0 & 0 & 0 & 0  \\
 0 & 0 & 0 & 0 & 0  \\
\end{array}
\right.$}} \nnb \\
&\nnb \\
&\hspace*{-47pt}
\scalebox{0.8}{\parbox{16cm}{$\dps\left.
\begin{array}{ccccccc}
 -L_2+L_3-L_4 & \frac{L_3}{4}-\frac{L_6}{4} & 0 & \frac{L_3}{4}-\frac{L_4}{4} & \frac{3 L_6}{2}-\frac{3 L_3}{2} &
   -\frac{L_3}{2}+\frac{L_4}{4}+\frac{L_6}{4} & \frac{L_6}{4}-\frac{L_3}{4} \\
 -L_2-L_4+\frac{L_5}{2} & \frac{L_5}{4} & \frac{3 L_5}{2} & \frac{L_4}{4}+\frac{L_5}{4} & 0 & -\frac{L_4}{4} & \frac{L_5}{4} \\
 0 & \frac{L_6}{2} & 0 & 0 & -2 L_6 & -\frac{L_6}{2} & -\frac{L_6}{2} \\
 0 & L_1-\frac{L_6}{2} & 0 & 0 & L_6 & \frac{L_6}{2} & \frac{L_6}{2} \\
 -L_5 & \frac{L_3}{2}-\frac{L_5}{2} & 2 L_3-2 L_5 & \frac{L_3}{2}-\frac{L_5}{2} & 0 & 0 & \frac{L_3}{2}-\frac{L_5}{2} \\
 -4 L_1+2 L_2-L_3-2 L_4+L_5 & -L_1-\frac{L_3}{2}+\frac{L_5}{2} & L_5-L_3 & \frac{L_5}{2}-\frac{L_3}{2} & 0 & 0 & \frac{L_5}{2}-\frac{L_3}{2} \\
 0 & -2 L_4 & 0 & 0 & 0 & 0 & L_3 \\
 0 & 0 & L_5 & L_1+L_4 & 0 & 0 & 0 \\
 0 & 0 & -6 L_5 & -4 L_1-4 L_4 & 0 & 0 & 0 \\
 0 & 0 & 0 & 0 & L_6-3 L_3 & L_1-L_3+L_4 & 0 \\
 0 & 0 & 0 & 0 & 6 L_3-6 L_6 & -4 L_1+2 L_3-4 L_4 & 0 \\
 0 & 0 & 0 & 0 & 0 & 0 & -L_3 \\
\end{array}
\right)  \, ,$}} \nnb \\ &\label{eq:Atilde112}
\end{align}
}

\vspace*{-25pt}

\allowdisplaybreaks{
\begin{align}
\hspace*{-2pt} \tilde A_{13-15} = &
\scalebox{0.8}{\parbox{16cm}{$\dps\left(
\begin{array}{ccccc}
 -2 L_4 & L_3-L_4 & L_2 & 0 & L_3-L_4 \\
 0 & 2 L_2-L_3+2 L_4-2 L_5 & -L_5 & \frac{L_3}{2}-\frac{L_5}{2} & 0 \\
 0 & -2 L_1-2 L_4+2 L_5 & -4 L_1+2 L_2-L_3-2 L_4+L_5 & -L_1-\frac{L_3}{2}+\frac{L_5}{2} & 0 \\
 0 & 0 & 0 & -2 L_4 & 0 \\
 0 & 0 & 0 & -\frac{L_3}{2} & 2 L_4-L_3 \\
 0 & 0 & 0 & 0 & 0 \\
 0 & 0 & 0 & 0 & 0 \\
 0 & 0 & 0 & 0 & 0 \\
 0 & 0 & 0 & 0 & 0 \\
\end{array}
\right.$}} \nnb \\
&\nnb \\
&\hspace*{202pt}
\scalebox{0.8}{\parbox{16cm}{$\dps\left.
\begin{array}{cccc}
 2 L_4-2 L_3 & \frac{L_4}{2}-\frac{L_3}{2} & 2 L_3-2 L_4 & 0 \\
 2 L_3-2 L_5 & \frac{L_3}{2}-\frac{L_5}{2} & 0 &
   \frac{L_3}{2}-\frac{L_5}{2} \\
 L_5-L_3 & \frac{L_5}{2}-\frac{L_3}{2} & 0 & \frac{L_5}{2}-\frac{L_3}{2} \\
 0 & 0 & 0 & L_3 \\
 0 & 0 & -2 L_3 & -\frac{L_3}{2} \\
 L_5 & L_1+L_4 & 0 & 0 \\
 -6 L_5 & -4 L_1-4 L_4 & 0 & 0 \\
 0 & 0 & 0 & 0 \\
 0 & 0 & 0 & -L_3 \\
\end{array}
\right)  \, ,$}} \label{eq:Atilde1315}
\end{align}
}

\vspace*{-20pt}

\allowdisplaybreaks{
\begin{align}
\hspace*{-2pt} \tilde A_{16-22} = &
\scalebox{0.8}{\parbox{16cm}{$\dps\left(
\begin{array}{ccc}
 2 L_2-2 L_3+2 L_4+L_5 & L_5-L_3 & -L_3+L_5+L_6  \\
 -2 L_2+2 L_3-2 L_4+L_5 & -2 L_2+L_3-2 L_4+L_5 & L_3+L_5-L_6  \\
 -2 L_1-3 L_2+2 L_3-5 L_4 & -L_1+L_3-L_4 & -2 L_1-2 L_2+L_3-4 L_4 \\
 0 & 0 & 0 \\
 0 & 0 & 0 \\
 0 & 0 & 0 \\
 0 & 0 & 0 \\
\end{array}
\right.$}} \nnb\\
&\nnb \\
&\hspace*{30pt}
\scalebox{0.8}{\parbox{16cm}{$\dps\left.
\begin{array}{cccc}
 2 L_5-2 L_3 & \frac{L_5}{2}-\frac{L_3}{2} & 0 & \frac{L_5}{2}-\frac{L_3}{2} \\
 L_3+2 L_5 & \frac{L_3}{2}+\frac{L_5}{2} & -L_5 & \frac{L_3}{2}+\frac{L_5}{2}\\
 -3 L_1-3 L_2+3 L_3-6 L_4 & -\frac{L_1}{2}-L_2+\frac{L_3}{2}-\frac{3 L_4}{2} & L_2+L_4 & -\frac{L_1}{2}-L_2+\frac{L_3}{2}-\frac{3 L_4}{2} \\
 L_3 & L_4 & 0 & 0 \\
 -6 L_3 & -4 L_4 & 0 & 0 \\
 0 & 0 & -2 L_1-2 L_4+L_5 & L_1+L_4 \\
 0 & 0 & 0 & 0 \\
\end{array}
\right)   \, ,$}} \label{eq:Atilde1622}
\end{align}
}
\allowdisplaybreaks{
\begin{align}
\hspace*{2pt} \tilde A_{23-27} = &
\scalebox{0.8}{\parbox{16cm}{$\dps\left(
\begin{array}{ccc}
 2 L_2-3 L_3+2 L_4+L_5 & L_5 & -L_3+L_5+L_6 \\
 -2 L_1-L_3-2 L_4+L_5 & -4 L_1+2 L_2-2 L_3-2 L_4+L_5 & -L_3+L_5-L_6 \\
 -3 L_1-2 L_2+3 L_3-5 L_4 & -L_2-L_4 & -4 L_1+L_3-4 L_4 \\
 0 & 0 & 0 \\
 0 & 0 & 0 \\
 0 & 0 & 0 \\
 0 & 0 & 0 \\
\end{array}
\right.$}} \nnb\\
&\nnb \\
&\hspace*{75pt}
\scalebox{0.8}{\parbox{16cm}{$\dps\left.
\begin{array}{cccc}
   \frac{L_3}{2}-\frac{L_5}{2} & 0 & 2 L_5-2 L_3 & \frac{L_5}{2}-\frac{L_3}{2} \\
   L_1+\frac{L_3}{2}-\frac{L_5}{2} & L_3-L_5 & 2 L_5-2 L_3 & \frac{L_5}{2}-\frac{L_3}{2} \\
   \frac{L_1}{2}-\frac{L_3}{2}+\frac{L_4}{2} & L_1-L_3+L_4 & -6 L_1+6 L_3-6 L_4 & -\frac{3 L_1}{2}+\frac{3 L_3}{2}-\frac{3 L_4}{2} \\
   -2 L_4 & 0 & 0 & L_3 \\
   0 & -2 L_1-L_3-2 L_4+L_5 & 0 & L_1+L_4 \\
   0 & 0 & -2 L_3 & 0 \\
   0 & 0 & 0 & -L_3 \\
\end{array}
\right) \, ,$}} \label{eq:Atilde2327}
\end{align}
}
\allowdisplaybreaks{
\begin{align}
\hspace*{-7pt} \tilde A_{28-32} = &
\scalebox{0.8}{\parbox{16cm}{$\dps\left(
\begin{array}{cccc}
 L_2-L_3-L_5 & L_2-2 L_3+2 L_4 & -L_1+L_2-L_3+L_5 & -\frac{L_1}{2}  \\
 L_2+2 L_4-L_5 & L_2 & L_1-L_2+L_5 & \frac{L_1}{2}  \\
 -L_2-3 L_5+2 L_7+2 L_8 & L_2-2 L_7-2 L_8 & -L_1+3 L_2-2 L_4+3 L_5-4 L_7-4 L_8 & -\frac{L_1}{2}  \\
 0 & 0 & 0 & -2 L_4 \\
 0 & 0 & 0 & 0  \\ 
 0 & 0 & 0 & 0 \\
 0 & 0 & 0 & 0 \\
 0 & 0 & 0 & 0 \\
 0 & 0 & 0 & 0 \\
 0 & 0 & 0 & 0 \\
\end{array}
\right.$}} \nnb\\
&\nnb \\
&\hspace*{-6pt}
\scalebox{0.8}{\parbox{16cm}{$\dps\left.
\begin{array}{cccccc}
   L_1+L_3-L_5 & -L_1-L_3+L_5 & \frac{L_1}{2} & 2 L_5-2 L_3 & 0 & -\frac{L_1}{2} \\
  -L_1 & L_1 & -\frac{L_1}{2} & 2 L_5-2 L_3 & 0 & \frac{L_1}{2} \\
  -L_1-L_3-L_5+2 L_7+2 L_8 & L_1+2 L_5-2 L_7-2 L_8 & -\frac{L_1}{2} & 6 L_5-6 L_7-6 L_8 & 2 L_8-2 L_7 & \frac{L_1}{2} \\
  0 & 0 & 0 & 0 & 0 & L_3 \\
  -2 L_1-L_3-2 L_4+L_5 & 0 & 0 & 0 & 0 & L_1+L_4 \\
  0 & -2 L_1-2 L_4+L_5 & L_1+L_4 & 0 & 0 & 0 \\
  0 & 0 & 0 & 0 & 0 & 0 \\
  0 & 0 & -\frac{L_4}{2} & L_3-3 L_4 & L_9 & \frac{L_4}{2} \\
  0 & 0 & -\frac{L_9}{2} & -3 L_9 & L_4-L_3 & \frac{L_9}{2} \\
  0 & 0 & 0 & 0 & 0 & -L_3 \\
\end{array}
\right) \, ,$}} \label{eq:Atilde2832}
\end{align}
}
\allowdisplaybreaks{
\begin{align}
\tilde A_{33,34} = &
\left(
\begin{array}{ccccccc}
 0 & 2 L_4 & 0 & 0 & -2 L_3 & 0 & 0 \\
 2 L_4-2 L_3 & -L_3 & \frac{L_3}{2} & -\frac{L_3}{2} & -2 L_3 & 0 & \frac{L_3}{2} \\
 0 & 0 & -2 L_4 & 0 & 0 & 0 & L_3 \\
 0 & 0 & 0 & 0 & 0 & 0 & 0 \\
 0 & 0 & 0 & -\frac{L_4}{2} & L_3-3 L_4 & L_9 & \frac{L_4}{2} \\
 0 & 0 & 0 & -\frac{L_9}{2} & -3 L_9 & L_4-L_3 & \frac{L_9}{2} \\
 0 & 0 & 0 & 0 & 0 & 0 & -L_3 \\
\end{array}
\right) , \label{eq:Atilde3334}\\
&\nnb \\
\tilde A_{35} = &
\left(
\begin{array}{ccc}
 -3 L_1-L_2-4 L_4+2 L_5 & L_3-L_5 & -\frac{L_1}{2} \\
 0 & L_3 & L_4 \\
 0 & -6 L_3 & -4 L_4 \\
\end{array}
\right) , \label{eq:Atilde35}\\
&\nnb \\
\tilde A_{36,37} = &
\left(
\begin{array}{cccc}
 2 L_5-2 L_1-2 L_2-4 L_4 & L_1+L_4 & L_2-L_3+L_4 & 0 \\
 0 & L_5-2 L_2-2 L_4 & 0 & L_2-L_3+L_4 \\
 0 & 0 & L_5-2 L_1-L_3-2 L_4 & L_1+L_4 \\
 0 & 0 & 0 & -L_3 \\
\end{array}
\right) \! . \label{eq:Atilde3637}
\end{align}}
%
%


\bibliographystyle{JHEP}

\providecommand{\href}[2]{#2}\begingroup\raggedright\endgroup

\end{document}